\documentclass[%
 preprint,%
 amssymb, amsmath,%
 aip,cha,%
]{revtex4-1}

\usepackage{docs}%
\usepackage{graphicx}
\usepackage{subfigure}
\usepackage{amsmath}
\usepackage{caption}
\usepackage{bm}%
\usepackage[colorlinks=true,linkcolor=blue]{hyperref}%
\usepackage{soul}
\soulregister\cite7
\soulregister\ref7
\soulregister\pageref7
\soulregister\label7

\usepackage{setspace}
\expandafter\ifx\csname package@font\endcsname\relax\else
 \expandafter\expandafter
 \expandafter\usepackage
 \expandafter\expandafter
 \expandafter{\csname package@font\endcsname}%
\fi
\graphicspath{{./Figures/}}

\begin{document}
\title{Mechanism of the fcc--to--hcp Phase Transformation in Solid Ar}%

\author{Bingxi Li}
\email[]{bxli@ucdavis.edu}
\affiliation{Department of Materials Science and Engineering, University of California, Davis, Davis, California 95616, USA}
\author{Guangrui Qian}
\affiliation{Department of Geosciences, Stony Brook University, Stony Brook, New York 11794, USA}
\author{Artem R. Oganov}
\affiliation{Department of Geosciences, Stony Brook University, Stony Brook, New York 11794, USA}
\affiliation{Skolkovo Institute of Science and Technology, Skolkovo Innovation Center, 5 Nobel St., Moscow 143026, Russia}
\affiliation{Department of Problems of Physics and Energetics, Moscow Institute of Physics
and Technology, Dolgoprudny City, Moscow Region 141700, Russia}
\affiliation{International Center for Materials Design, Northwestern Polytechnical University, Xi'an, 710072, China}
\author{Salah Eddine Boulfelfel}
\affiliation{School of Chemical and Biomolecular Engineering, Georgia Institute of Technology, Atlanta, Georgia 30332, USA}
\author{Roland Faller}
\email[]{rfaller@ucdavis.edu}
\affiliation{Department of Chemical Engineering, University of California, Davis, Davis, California 95616, USA}

\date{\today}%

\begin{abstract}
We present an atomistic description of the {\it fcc}--to--{\it hcp} transformation mechanism in solid argon (Ar) obtained from transition path sampling molecular dynamics simulation. The phase transition pathways collected during the sampling for an 8000--particle system reveal three transition types according to the lattice deformation and relaxation details. In all three transition types, we see a critical accumulation of defects and uniform growth of a less ordered transition state, followed by a homogeneous growth of an ordered phase. Stacking disorder is discussed to describe the transition process and the cooperative motions of atoms in \{111\} planes. We investigate the nucleation with larger system. In a system of 18000--particles, the collective movements of atoms required for this transition are facilitated by the formation and growth of stacking faults. However the enthalpy barrier is still far beyond the thermal fluctuation. The high barrier explains previous experimental observations of the inaccessibility of the bulk transition at low pressure and its sluggishness even at extremely high pressure. The transition mechanism in bulk Ar is different from Ar nanoclusters as the orthorhombic intermediate structure proposed for the latter is not observed in any of our simulations.
\end{abstract}
\maketitle


\section{Introduction}

Rare gas solids (RGSs) of neon, argon, krypton and xenon crystallize in face--centered cubic ({\it fcc}) structures at ambient pressure and low temperatures~\cite{barron1955cubic,jansen1963absolute}. However, early experimental studies and theoretical predictions pointed out the possibility of a hexagonally close--packed ({\it hcp}) structure~\cite{meyer1964new,barrett1964x,schuberth1976evidence,SONNENBLICK1977276,PhysRevLett.67.3263}. The {\it hcp} structure can coexist with {\it fcc} as a metastable phase in pure rare gas solids at low temperatures~\cite{meyer1964new,barrett1964x,schuberth1976evidence,SONNENBLICK1977276} and becomes stable in solid solutions~\cite{meyer1964new,gal2004influence,curzon1971electron}. Calculations based on two-- and many--body interaction potentials have predicted the {\it hcp} structure to be energetically more favorable than the {\it fcc} polymorph~\cite{schwerdtfeger2006extension,krainyukova2011role}. This stability order is reversed after including zero--point vibrational effects~\cite{schwerdtfeger2006extension}.

Even though the stability of {\it fcc} and {\it hcp} phases under ambient pressure has been disputed for decades, stability of {\it hcp} structures is evidenced by many high--pressure experiments~\cite{errandonea2002crystal,errandonea2002phase,errandonea2006structural,jephcoat1987pressure,PhysRevB.77.052101,freiman2008raman,PhysRevB.79.132101}. Rare gas solids martensitically transform from {\it fcc} to {\it hcp} before metallization occurs under pressure~\cite{errandonea2002crystal,errandonea2002phase,errandonea2006structural,jephcoat1987pressure}. However, the pressure--induced {\it fcc}--to--{\it hcp} transition in RGSs is rather sluggish and the two phases coexist over a wide range of pressures. X--ray diffraction and Raman spectroscopy show a transformation pressure in the range of 1.5--41 and 3.2--50~GPa for Xe and Kr, respectively, and beyond 49.6~GPa for Ar~\cite{errandonea2002crystal,errandonea2002phase,errandonea2006structural,PhysRevB.77.052101,freiman2008raman,PhysRevB.79.132101}. Theoretically, sluggishness of the {\it fcc} to {\it hcp} transformation in Xe was attributed to a high energy barrier~\cite{kim2006martensitic}. A first--principles study determined the enthalpy barrier for a stacking disorder growth pathway~\cite{cynn2001martensitic} at lower pressure and an alternative pathway involving an orthorhombic distortion at higher pressure.

Recent theoretical and experimental studies~\cite{danylchenko2004electron,danil2008electron,danylchenko2014electron} investigated the phase behavior of Ar clusters under ambient pressure. It was shown that increasing the Ar cluster size led to a transition from {\it fcc} to {\it fcc}/{\it hcp} mixed structures during cluster growth. An orthorhombic structure was predicted as intermediate in the {\it fcc}--to--{\it hcp} transition as it accounts for the diffraction peaks originating from neither in {\it fcc} nor {\it hcp} structures in Ar nanoclusters~\cite{verkhovtseva2003atomic, krainyukova2012observation}. Although the {\it fcc}--to--{\it hcp} transition mechanism was well studied for rare gas clusters, a comprehensive understanding of this transformation in the corresponding bulk materials is still elusive.

Molecular dynamics is a method for mechanistic investigations. However, the enthalpy barrier of transition greatly reduces the efficiency of finding the transition path. The transition path sampling~\cite{bolhuis2002transition,dellago2006transition} method is designed to solve this problem. Therefore, we employ transition path sampling to provide detailed atomistic understanding of the transition mechanism and the ensuing structures.

\section{Transition Path Sampling molecular dynamics}
This section briefly reviews transition path sampling (TPS)~\cite{bolhuis2002transition,dellago2006transition} for transition pathways with deterministic dynamics~\cite{A801266K}. Besides, we present the identification of the initial and final state of transition pathways using a fingerprint function~\cite{oganov2009quantify}. The characterization of structures formed during transition is also discussed.

\subsection{Transition Path Sampling}
TPS performs importance sampling in a transition path ensemble with two main trial moves, shooting and shifting.

A shooting move randomly chooses a time slice of a trajectory and adds a perturbation $\delta p$ drawn from a Gaussian distribution to its atomic momenta. The selected time slice $x^{(o)}_{t}$ at time $t$ on the old path will yield a new phase space point $x^{(n)}_{t}$ as the starting point for a new trajectory. A new trajectory with time length being $T$ is generated by shooting forward to time $T$ and backward to time $0$ from time $t$. For deterministic dynamics, the acceptance probability of this move is:
\begin{equation}
P_{acc}^{o\rightarrow n} = h_A[x_0^{(n)}] h_B[x_T^{(n)}] min\bigg[1,\frac{\rho(x_{t}^{(n)})}{\rho(x_{t}^{(o)})}\bigg]
\end{equation}
where $\rho(x_{t}^{(o)})$ and $\rho(x_{t}^{(n)})$ are the equilibrium phase space distribution of the old and new time slices at time $t$, respectively. $h_A[x_0^{(n)}]$ has a binary value: 1 for the beginning structure of the new trajectory, $x_0^{(n)}$ in state A and 0 otherwise. $h_B[x_T^{(n)}]$ is defined equivalently for the ending structure of the new trajectory, $x_T^{(n)}$.

A shifting move translates the old trajectory by a time shift $\Delta t$ that can be drawn from a Gaussian distribution. The corresponding acceptance probability $P_{acc}^{o\rightarrow n}$ is
\begin{equation}
P_{acc}^{o\rightarrow n} = h_A[x_0^{(n)}] h_B[x_T^{(n)}]
\end{equation}

The initial trajectory connecting the initial and final state which is needed to start transition path sampling simulations was generated by the variable--cell nudged elastic band~\cite{qian2013variable} method. For the following transition path sampling molecular dynamics simulations we used the TPS module from the USPEX package~\cite{glass2006uspex,oganov2011evolutionary,lyakhov2013new,boulfelfel2012understanding}. In the TPS simulation of the system of 8000 Ar atoms, 7427 distinct trajectories were created by 17334 shooting and shifting moves during the sampling. In a larger system of 18000 Ar atoms, 8688 trajectories were sampled after 20410 shooting and shifting moves.

\subsection{Molecular Dynamics Simulation}
Classical molecular dynamics simulations are carried out using the LAMMPS code~\cite{Plimpton19951} to generate the new trajectories after each shooting or shifting move. The velocity--Verlet algorithm with an integration time step of 0.1~fs is used to ensure time--reversibility. The simulation employed a Nose--Hoover thermostat~\cite{PhysRevA.31.1695} with a relaxation time of 30~fs and a Nose--Hoover barostat~\cite{PhysRevA.34.2499} with a relaxation time of 300~fs, ensuring an $NpT$ ensemble at $T = 40$~K and $p = 1$~bar~\cite{krainyukova2012observation}. The simulation box containing 8000 or 18000~Ar atoms allows anisotropic shape changes to avoid biasing the evolution of the dynamics and the resulting transition mechanism.

The interatomic interactions were modeled using a Lennard--Jones (LJ) 12--6 model with Ar interaction parameters $\sigma=3.405$~\AA\ and $\epsilon=0.238$~kcal/mol~\cite{hirschfelder1964theory,schwerdtfeger2006extension}, and truncated at 10~\AA. The LJ potential has been widely applied to study rare gas solids, e.g. their melting~\cite{schwerdtfeger2006extension,PhysRevB.71.134106}. 

\subsection{Characterization of Structures}
Structures encountered during the phase transition usually have low or even no crystallinity. To characterize these transitory structures, we introduce the structure similarity function and coordination numbers to analyze their lattice deformation and the local packing.

The structure similarity value is used as the order parameter to measure the phase transition boundary. It refers to the cosine similarity between the fingerprint vectors of different structures~\cite{oganov2009quantify}. Each structure is represented by a fingerprint vector generated by the values of its fingerprint function $f_x(r)$ in each bin (of width $\Delta$).

The fingerprint function $f(r)$ subtracts 1 from the radial distribution function (RDF) and is short--ranged. The fingerprint vector $F$ is obtained through its discretization. Each structure is therefore uniquely described as a vector in the fingerprint space, the dimensionality of which equals the number of discretization bins. For a structure $x$ during the phase transition, the fingerprint vector $F_x$ can be written as,
\begin{equation}
  \begin{split}
    F_x & = [f_{x}(r_0), f_{x}(r_1),\dots, f_x(r_n)]\\
        & = [g_x(r_0)-1,\ g_x(r_1)-1,\ \dots\ ,\ g_x(r_n)-1]
  \end{split}
\end{equation}
where $g_x(r)$ is the RDF value of structure $x$ at distance $r$. Our structure similarity function measures similarity between structures by calculating the cosine similarity in fingerprint space. The similarity of structure $x$ to the {\it fcc} phase is then determined using the structure similarity function $S_{fcc}(x)$,
\begin{equation}
S_{fcc}(x)=\frac{\overrightarrow{F}_x\cdot \overrightarrow{F}_{fcc}}{|\overrightarrow{F}_x||\overrightarrow{F}_{fcc}|}
\end{equation}
which is the cosine similarity between the fingerprint vector of structure $x$ and the {\it fcc} phase. The structure similarity function ranges from $0$ meaning totally different, to $1$ meaning identical; in--between values indicate intermediate similarity. In this paper, the {\it fcc}--structure similarity criterion is set to be $0.999$, i.e. if $S_{fcc}\geq0.999$, we identify a structure as {\it fcc}. The similarity to the {\it hcp} phase is defined analogously. Structures with $S_{fcc}<0.999$ and $S_{hcp}<0.999$ are regarded as transitory structures. Transitory structures define the transition region for each trajectory.

By comparing the percentages of differently coordinated atoms in each transitory structures, we can develop a view into the local packing changes during phase transition. The coordination number is computed as the number of neighbor atoms within the specified cutoff distance from the central atom. The cutoff distance is placed on the first local minimum of the RDF of each structure. The cutoff varies from structure to structure along the trajectory. The first minimum is at the shortest distance for crystalline structures and at larger distances for intermediate structures. The first local minimum is the same for the two close--packed structures. In {\it fcc} and {\it hcp} structures, the coordination number of each atom is~12, equal to the number of nearest neighbors. Over-- and under--coordinated atoms appear during the transition.

\section{Results}
\subsection{Small System}
We first discuss the 8000--particle system. Through transition path sampling molecular dynamics simulation, we developed atomistic understanding of {\it fcc}--to--{\it hcp} phase transformation in Ar solid at 40~K under ambient pressure. The transition pathways are categorized into three types according to the lattice deformation and relaxation details. But all three types of transitions go through the same transition state with a high enthalpy barrier. The enthalpy barrier is $578\pm 50$~kcal/mol for the {\it fcc}$\rightarrow${\it hcp} transition in Ar solid modeled by a system of 8000 atoms, which is far beyond the thermal energy at 40~K ($kT=0.0795$~kcal/mol).

\begin{figure}[tbp!]
\captionsetup{justification=centering}
\includegraphics[width=0.5\textwidth]{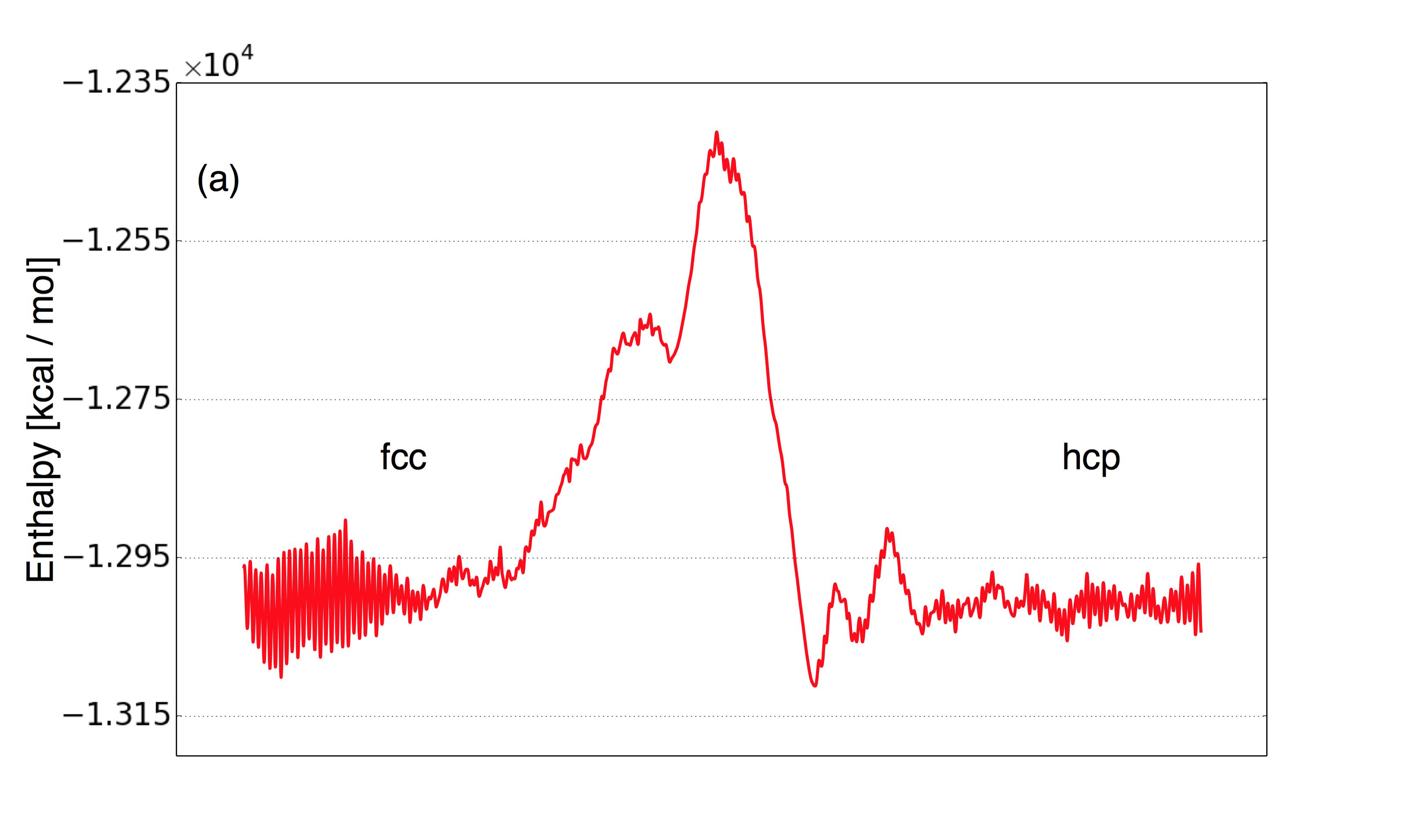}
\includegraphics[width=0.5\textwidth]{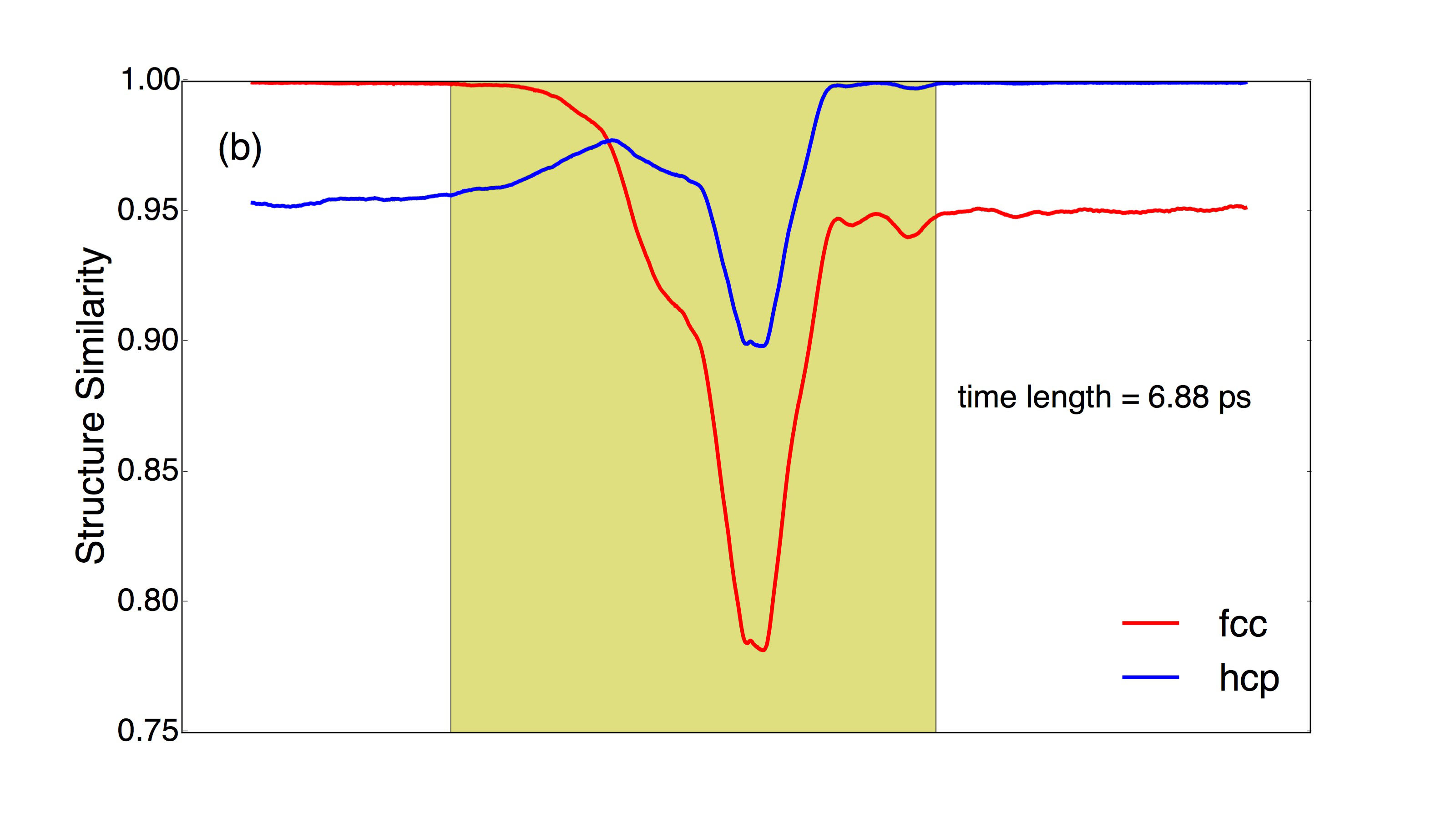}
\includegraphics[width=0.5\textwidth]{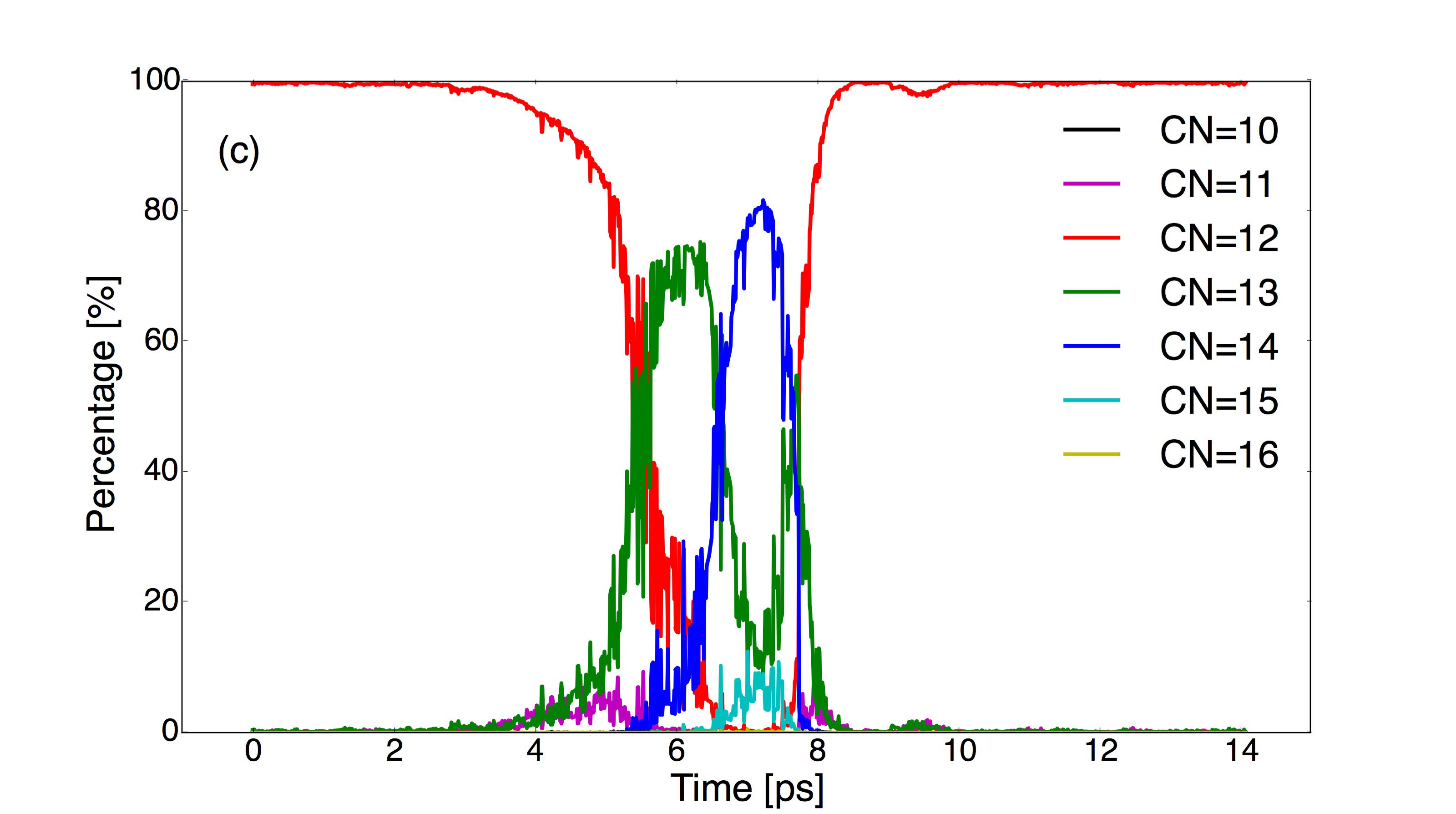}
\caption{\label{fig:11899-tsc}(a) Enthalpy of trajectory I. (b) The transition region (yellow) of trajectory I. Structure similarities to {\it fcc} and {\it hcp} phases are plotted in red and blue. (c) The percentages of atoms with different coordination numbers along trajectory I.}
\end{figure}

\subsubsection{Three Types of Transition Pathways}

\begin{figure}[bp!]
\captionsetup{justification=centering}
\includegraphics[width=0.5\textwidth]{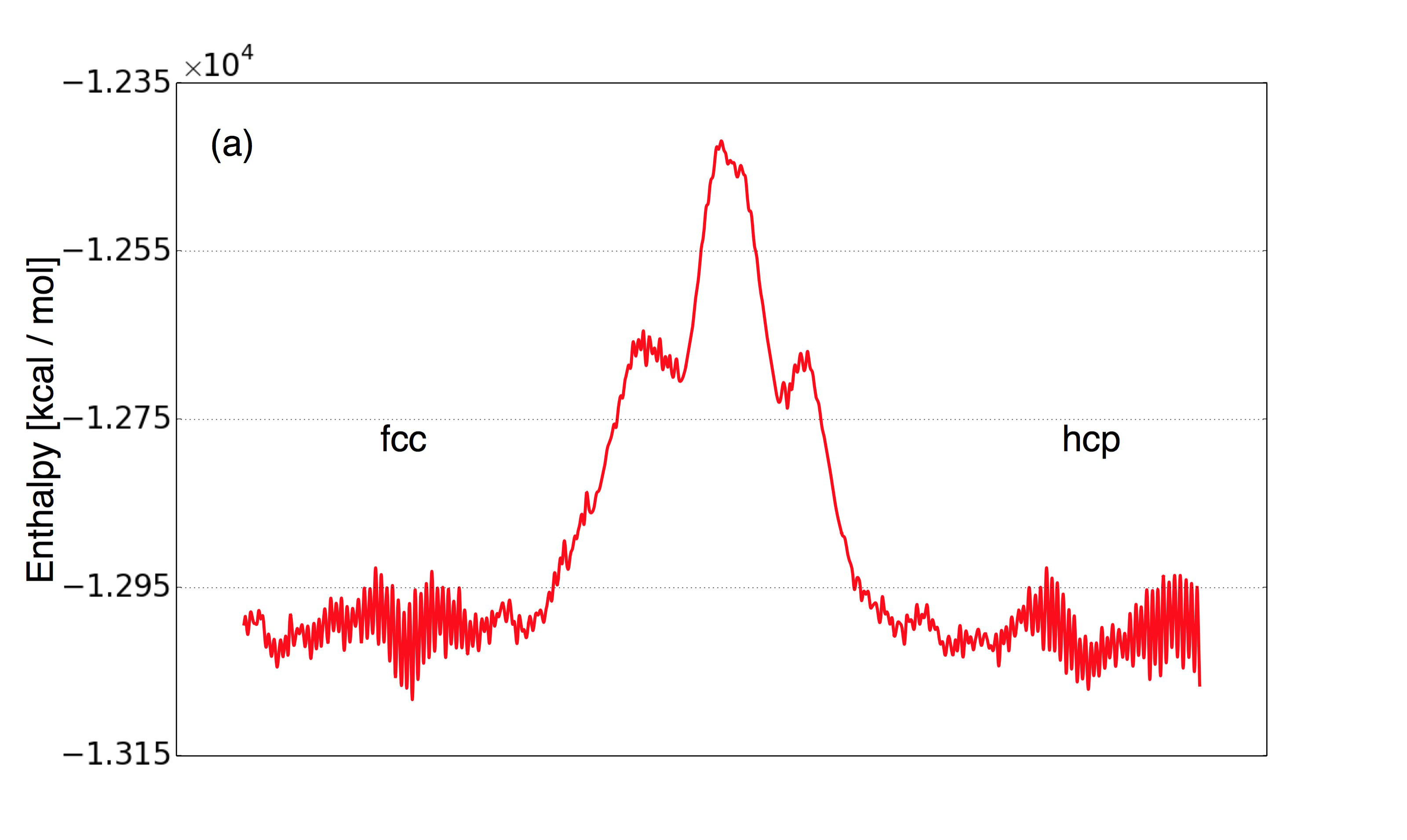}
\includegraphics[width=0.5\textwidth]{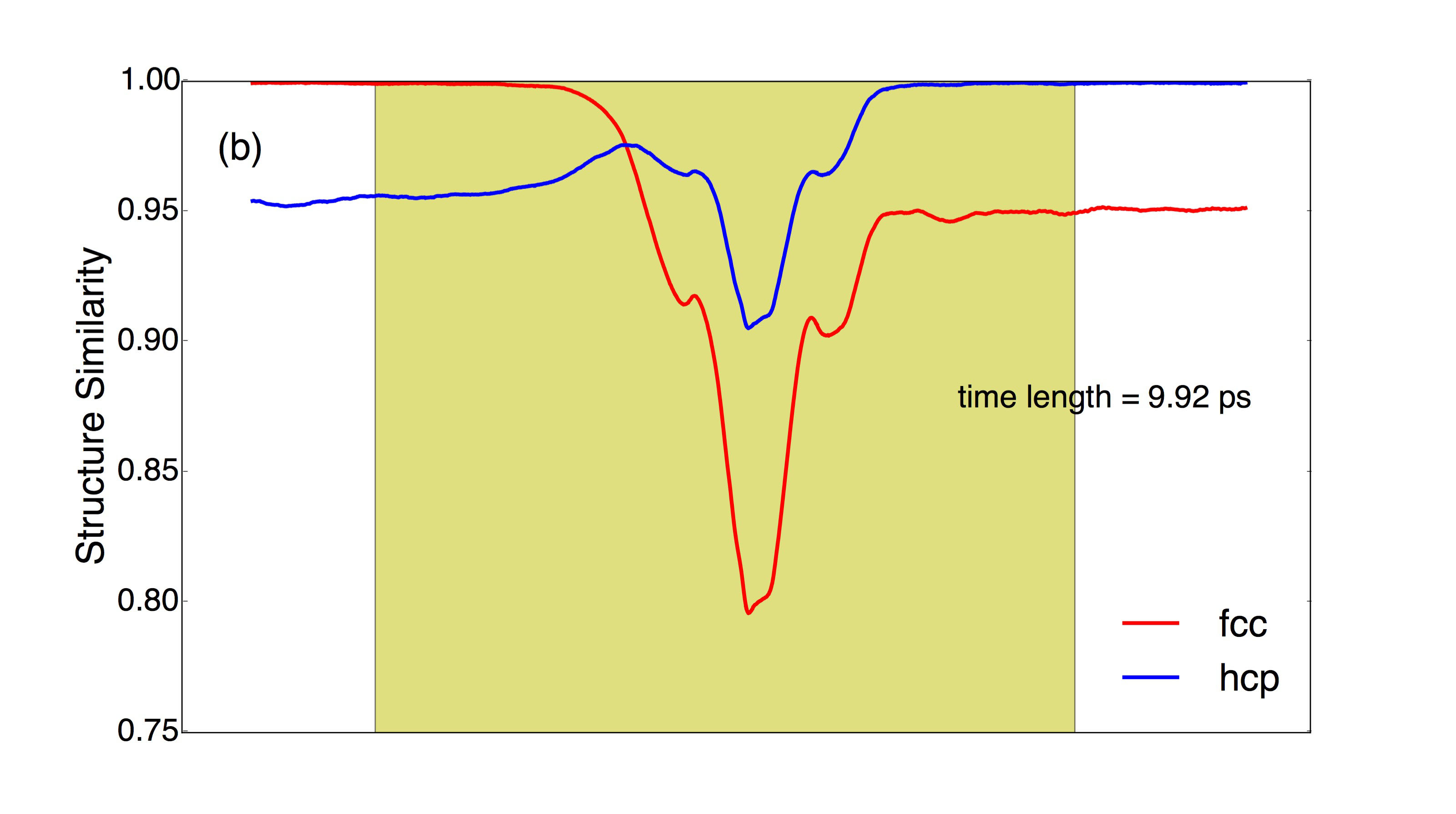}
\includegraphics[width=0.5\textwidth]{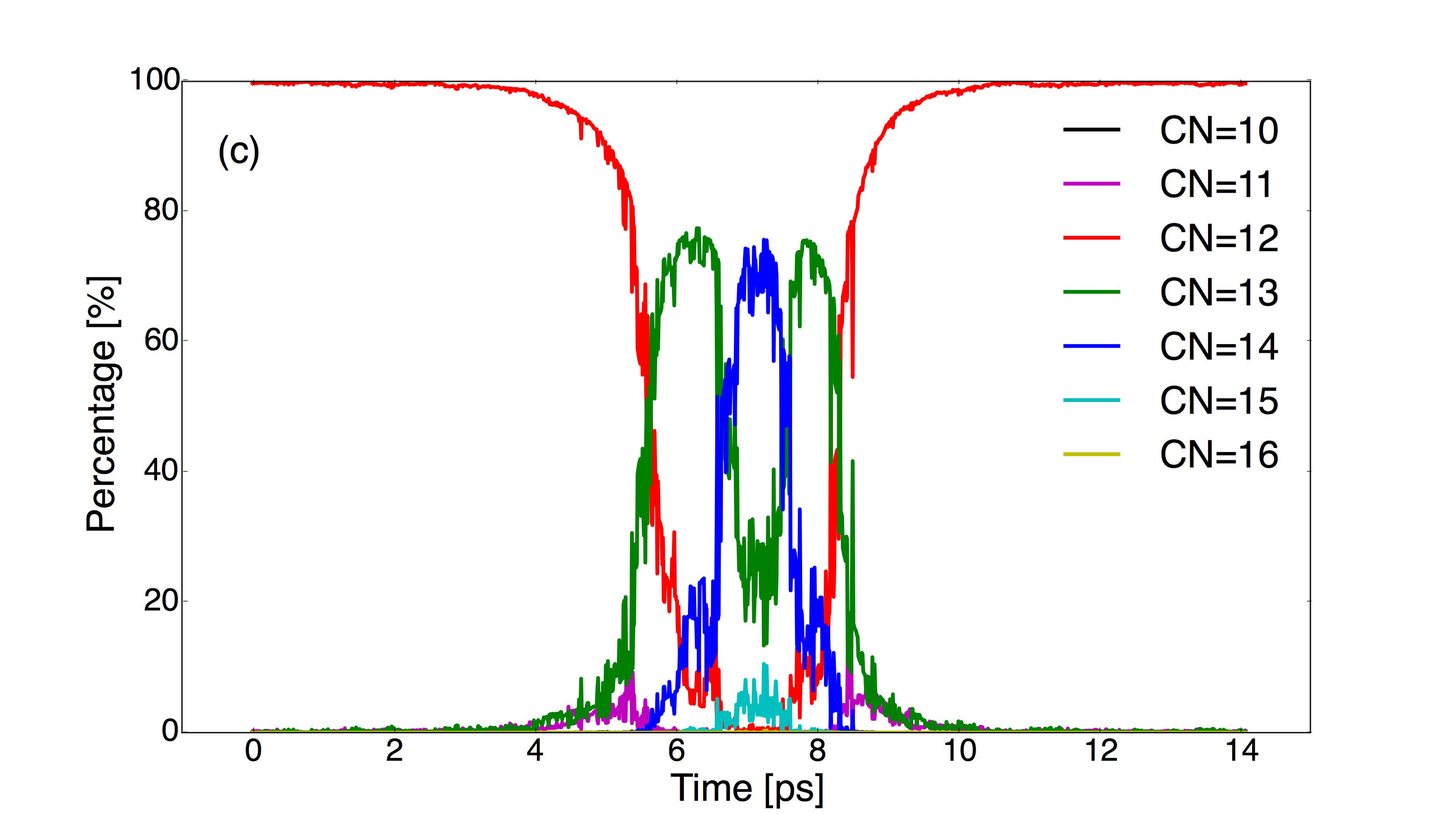}
\caption{\label{fig:13494-tsc}(a) Enthalpy of trajectory II. (b) The transition region (yellow) of trajectory II. Structure similarities to {\it fcc} and {\it hcp} phases are plotted in red and blue. (c) The percentages of atoms with different coordination numbers along trajectory II.}
\end{figure}

\begin{figure}[bp!]
\captionsetup{justification=centering}
\includegraphics[width=0.5\textwidth]{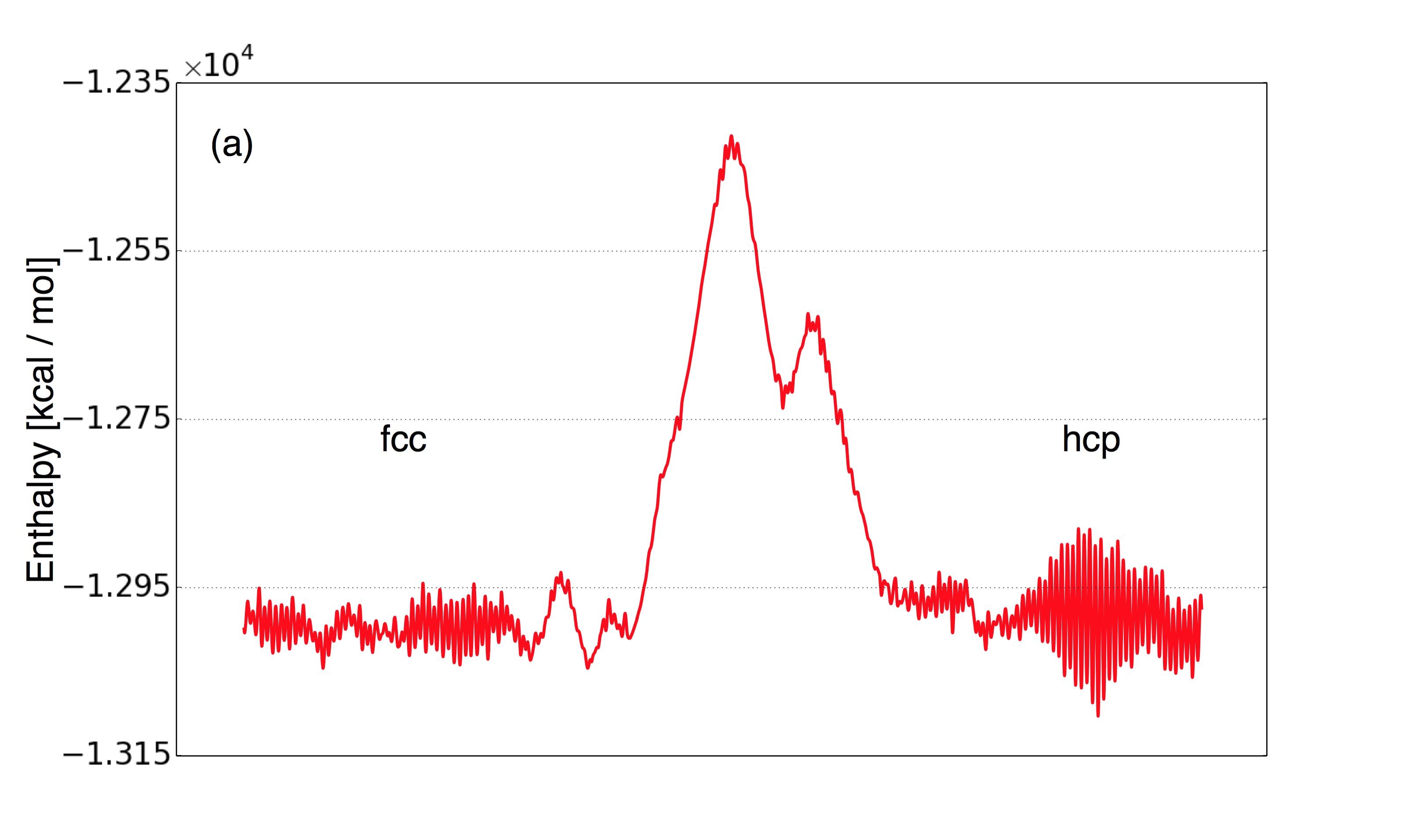}
\includegraphics[width=0.5\textwidth]{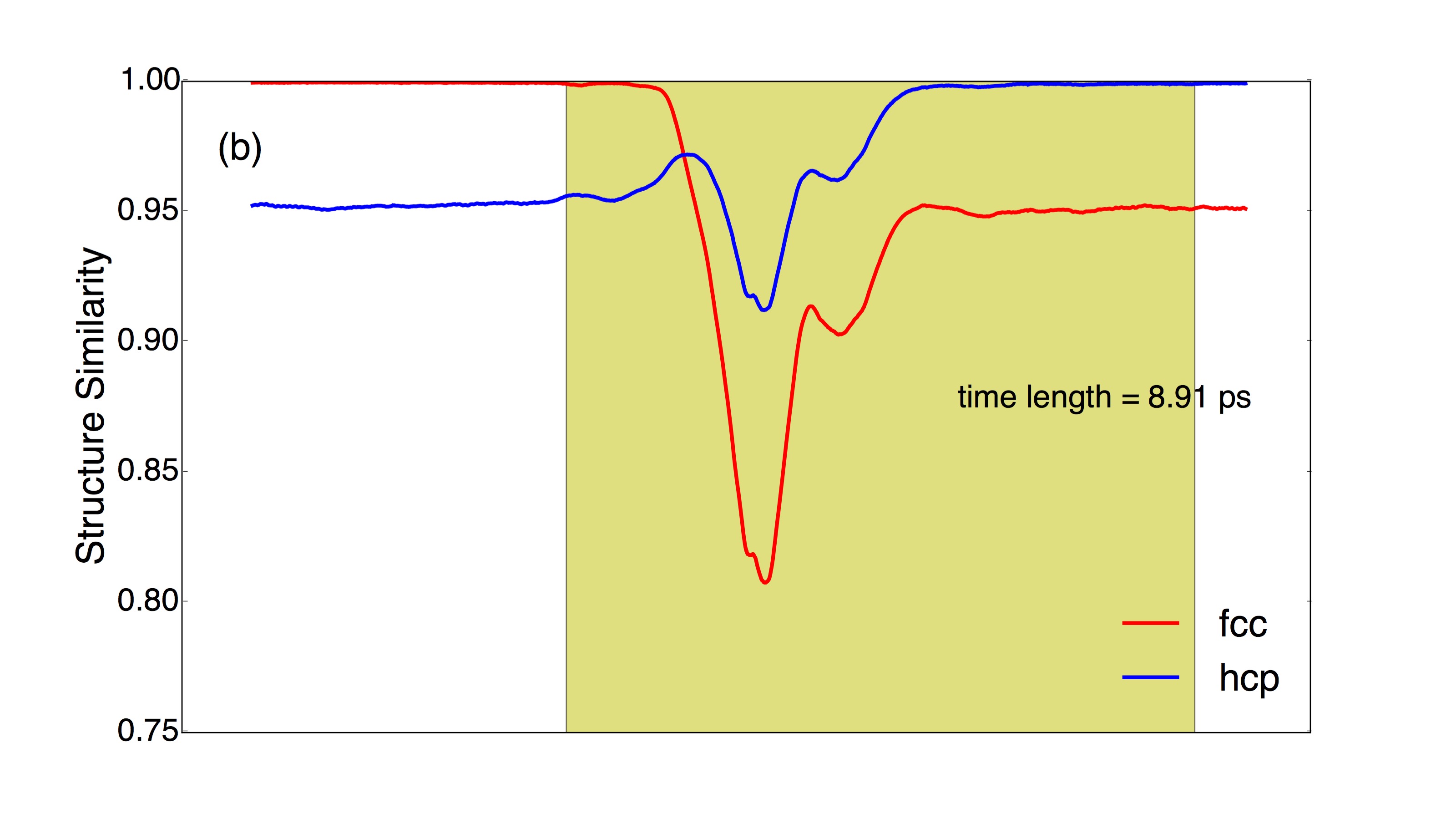}
\includegraphics[width=0.5\textwidth]{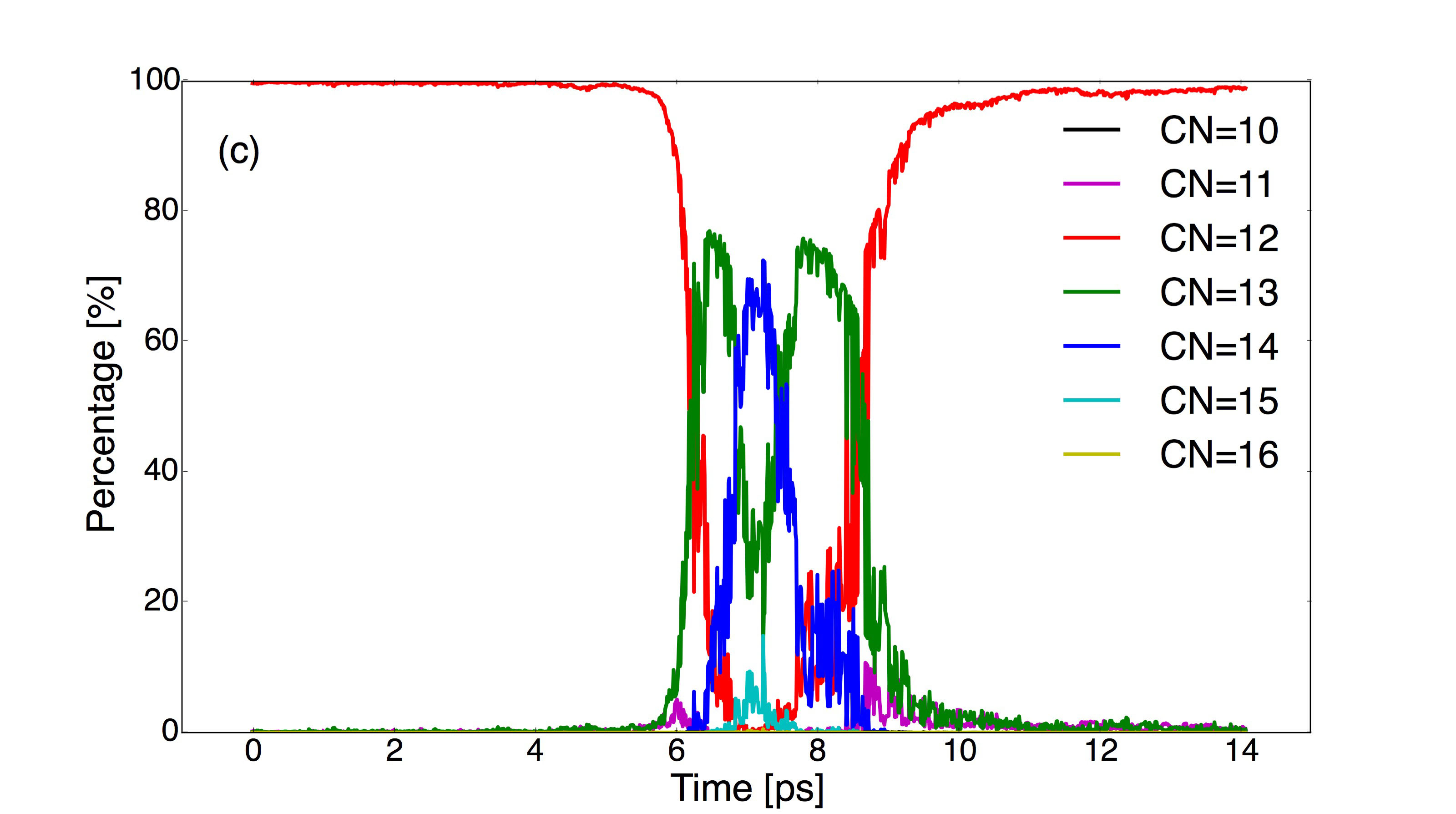}
\caption{\label{fig:16350-tsc}(a) Enthalpy of trajectory III. (b) The transition region (yellow) of trajectory III. Structure similarities to {\it fcc} and {\it hcp} phases are plotted in red and blue. (c) The percentages of atoms with different coordination numbers along trajectory III.}
\end{figure}

We are defining 3 types of transitions. Trajectory \uppercase\expandafter{\romannumeral1} represents the Type \uppercase\expandafter{\romannumeral1} transition. The structure deforms slowly from the {\it fcc} phase to the transition state but relaxes quickly to the {\it hcp} phase. The growth and decay rates of local defects dominate the transformation process. Fig.~\ref{fig:11899-tsc}(b) and (c) indicate the consistency between the lattice deformation and local defects. When more than $1.5\%$ of atoms become over-- or under--coordinated the lattice looses its similarity to {\it fcc} at $2.81$~ps. According to Fig.~\ref{fig:11899-tsc}(a) and (c), the increasing enthalpy follows the local packing change from the beginning of transition. When the system reaches the local minimum in enthalpy at $6.25$~ps over $80\%$ atoms are not 12--coordinated. After waiting for sufficient thermal fluctuations to activate further local packing changes, more local defects are generated to accelerate the lattice deformation. When the system has the highest enthalpy and largest number of local defects at $6.94$~ps, the structure has least similarity to both {\it fcc} and {\it hcp} phase. $99.83$\% atoms become over--coordinated. The transition intermediate is in neither of the two crystal phases. More details on the crystallinity of intermediate states are presented in the supplementary material. The lattice afterwards quickly relaxes to the {\it hcp} phase. Although the average arrangement of atoms first demonstrates the character of {\it hcp} at $8.67$~ps, the coordination numbers still stabilizes to 12 until $9.69$~ps in order to dissipate the energy fluctuations.

\begin{figure*}[htbp!]
\captionsetup{justification=centering}
\includegraphics[width=\linewidth]{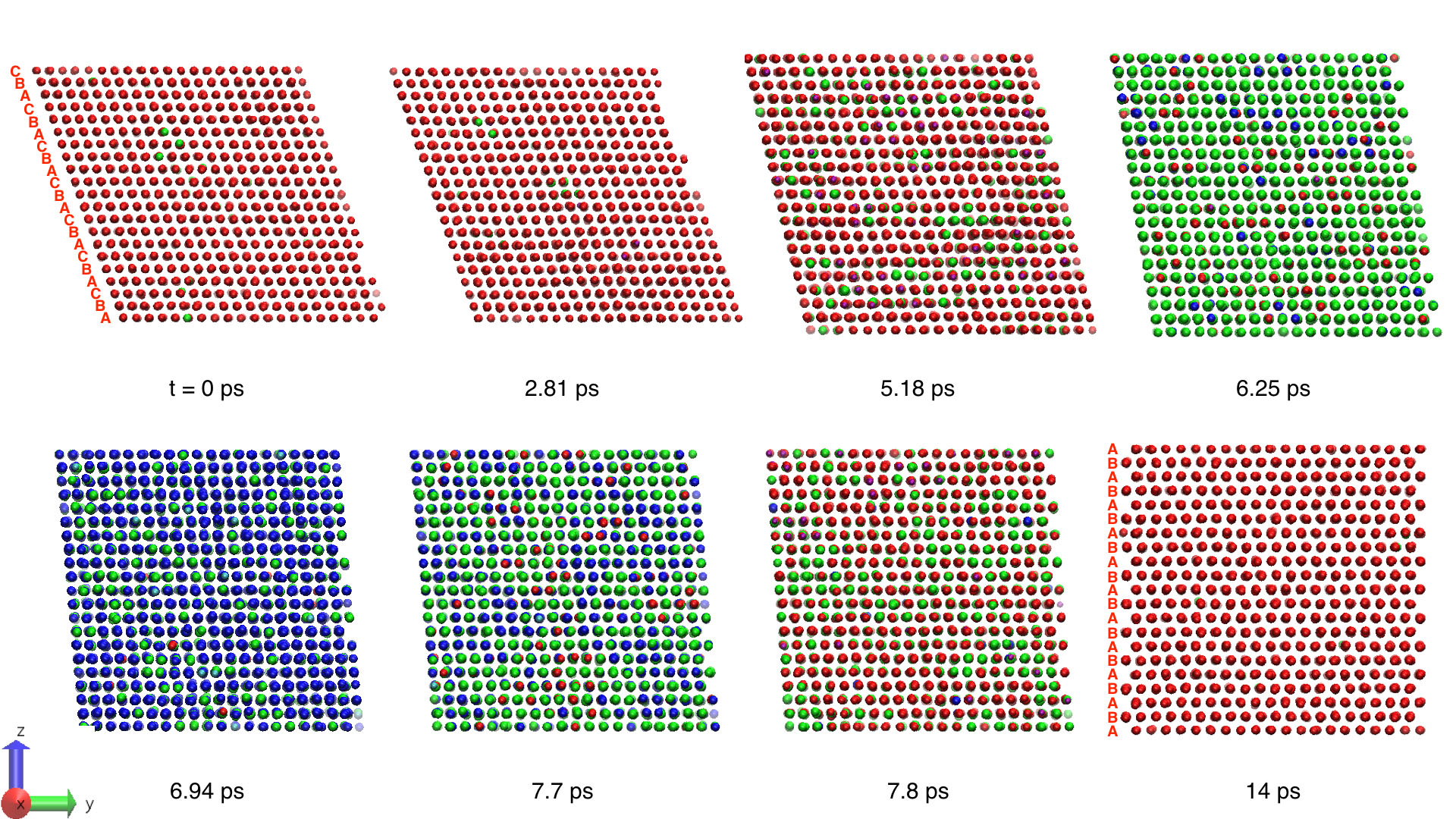}
\caption{\label{fig:transition}Snapshots of configurations during a transition in 8000--particle system, different colors are different coordination numbers. Atoms in red are 12--coordinated while green and blue are for 13-- and 14--coordinated atoms.}
\end{figure*}

A Type \uppercase\expandafter{\romannumeral2} transition can be observed in trajectory II (FIG.~\ref{fig:13494-tsc}). During the transition, the lattice deformation and relaxation take almost equal time. Based on the structure similarity, the lattice begins to deform at $1.74$~ps and reaches the largest deformation $5.29$~ps later. Then it takes $4.63$~ps to relax into the {\it hcp} phase. As shown in Fig.~\ref{fig:13494-tsc}(a), two metastable states appear during the lattice deformation and relaxation. The first metastable state forms at $6.39$~ps and soon accumulates thermal energy to produce more local defects, which facilitate the following lattice deformation. After overcoming the transition saddle at $7.03$~ps, the system is trapped into a second metastable state. It just takes 0.39~ps to overcome the small enthalpy barrier and finally relax to the {\it hcp} phase at $11.66$~ps.

Fig.~\ref{fig:16350-tsc} indicates a Type \uppercase\expandafter{\romannumeral3} transition, in which the lattice experiences a fast deformation and slow relaxation. Starting with a small thermal fluctuation at $4.45$~ps, only $1.5$\% of atoms become not--12--coordinated until $5.56$~ps. The structure still keeps $99.8$\% similarity to {\it fcc}. However, induced by a large fluctuation at that point, the lattice experiences a rapid change. $99.87$\% of atoms become over--coordinated within $1.7$~ps. As a result, the transition state has the largest deformation from both {\it fcc} and {\it hcp} phases at $7.26$~ps. A following increase of 12--coordinated atoms reduces the local defects and leads to a metastable phase with more structural similarity to {\it hcp} phase at $7.91$~ps. Another 0.51~ps is required to overcome the enthalpy barrier. The relaxation to {\it hcp} is completed at $13.36$~ps. The whole transition as characterized by structure similarity takes $8.91$~ps.

Looking at the enthalpy profiles along the three transformation paths, we see that their activation enthalpies are very close to each other. From an enthalpy viewpoint alone one could conclude that these paths have equal chances to occur, but this would ignore entropy effects. A simple way to judge the likelihood of occurrence of each path, taking entropy into account is to count the frequency of occurrence of each type of pathway. We find that the percentage of Type \uppercase\expandafter{\romannumeral1} is 25.6\%, Type \uppercase\expandafter{\romannumeral2} is 13.0\%, and Type \uppercase\expandafter{\romannumeral3} is 61.4\%. We therefore assume that Type \uppercase\expandafter{\romannumeral3} is the most likely but the other ones have a significant share.

Despite the difference in lattice deformation and relaxation, the three types of transition all go through less ordered intermediates at this small system size. The order of the intermediate state is between fully amorphous and a crystalline solid. These intermediate structures are dominated by over--coordinated atoms, most of which are 13-- and 14--coordinated. It is noticeable that the intermediate structure in Type \uppercase\expandafter{\romannumeral3} transition has the highest similarity to {\it fcc} and {\it hcp} phases. Therefore less structural deformation is required in this type of transition.

\subsubsection{Transition Mechanism}
\begin{figure}[htbp!]
\captionsetup{justification=centering}
\includegraphics[width=0.5\textwidth]{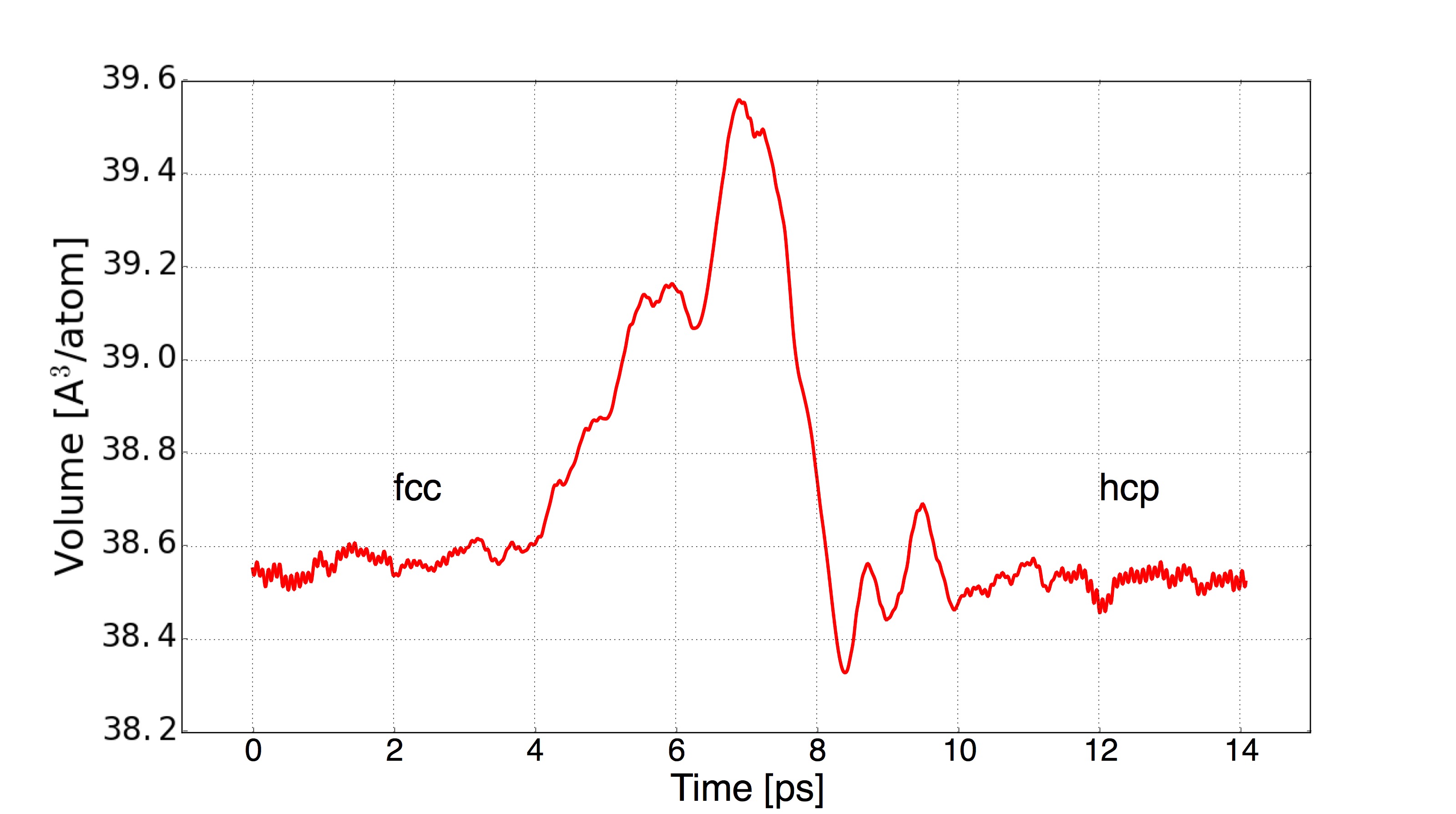}
\caption{\label{fig:11899-volume}Volume change in a Type I trajectory.}
\end{figure}

In FIG.~\ref{fig:transition}, we observe the phase growth along trajectory~I. At the beginning, nearly all atoms in the {\it fcc} structure are 12--coordinated. The lattice is characterized by a stacking sequence \ldots{\it ABCABC}\ldots. More than $80$\% atoms remain 12--coordinated since the lattice starts to deform at $2.81$~ps. Then some 13--coordinated atoms appear and distribute uniformly throughout the lattice. These 13--coordinated atoms form a metastable phase at $6.25$~ps and further transform into another intermediate, which is mainly dominated by 14--coordinated atoms. At $6.94$~ps, $99.83$\% of atoms become 14--coordinated. When the lattice begins to relax from the largest deformation as shown in FIG.~\ref{fig:11899-tsc}(b), 12--coordinated atoms grow homogeneously within the cell. The lattice finally relaxes to {\it hcp} by $8.67$~ps. The whole system ends with the {\it hcp} phase at $14$~ps with a stacking sequence of \ldots {\it ABABAB}\ldots. Fig~\ref{fig:11899-volume} illustrates details of the volume change of the cell shown in Fig~\ref{fig:transition}. There is $2.59$\% expansion of the cell during the phase transition. The consistencies of volume change with enthalpy, structure similarity and coordinates are explained in supplementary material.

Fig.~\ref{fig:transition} shows the homogeneous growth of the intermediate and the new ordered phase during the {\it fcc}--to--{\it hcp} transition in Ar solid. Through the collective sliding of planes, accompanied by the formation of defects, stacking sequence changes from \ldots{\it ABCABC}\ldots in {\it fcc} to \ldots{\it ABABAB}\ldots in {\it hcp}. The cooperative movement of many atoms not only results in the {\it fcc} stacking growth into {\it hcp} domains but also leads to the lattice deformation. It hereby explains the consistency between lattice deformation and local packing shown in Fig.~\ref{fig:11899-tsc}. The stacking disorder growth mechanism is also observed in another two types of phase transition. This mechanism was previously observed in experimental studies of {\it fcc}--to--{\it hcp} transition in Xe and Kr~\cite{cynn2001martensitic,errandonea2002phase}. A first--principles calculation suggested this mechanism as the transition pathway in Xe at lower pressure~\cite{kim2006martensitic}. An X--ray diffraction study on solid Ar up to 114 GPa predicted the development of stacking disorder during the transition~\cite{errandonea2006structural}. However, there is still not enough evidence to prove this assumption under ambient pressure as the {\it fcc}--to--{\it hcp} transition in Ar requires strong compression above 49.6 GPa~\cite{errandonea2002crystal,errandonea2002phase,errandonea2006structural,PhysRevB.77.052101,freiman2008raman,PhysRevB.79.132101}. Our previous results on the high enthalpy barrier explain the inaccessibility of this transition in bulk at low pressure. A recent study on {\it fcc}--to--{\it hcp} transition in Ar cluster at ambient pressure observes an orthorhombic intermediate\cite{krainyukova2012observation} while our stacking disorder growth mechanism in bulk Ar suggests an less ordered intermediate mainly due to the high appearance of defects in parallel to the sliding. The system generates such a high density of defects throughout that it appears no longer crystalline.

\begin{figure}[htbp!]
\captionsetup{justification=centering}
\includegraphics[width=0.5\textwidth]{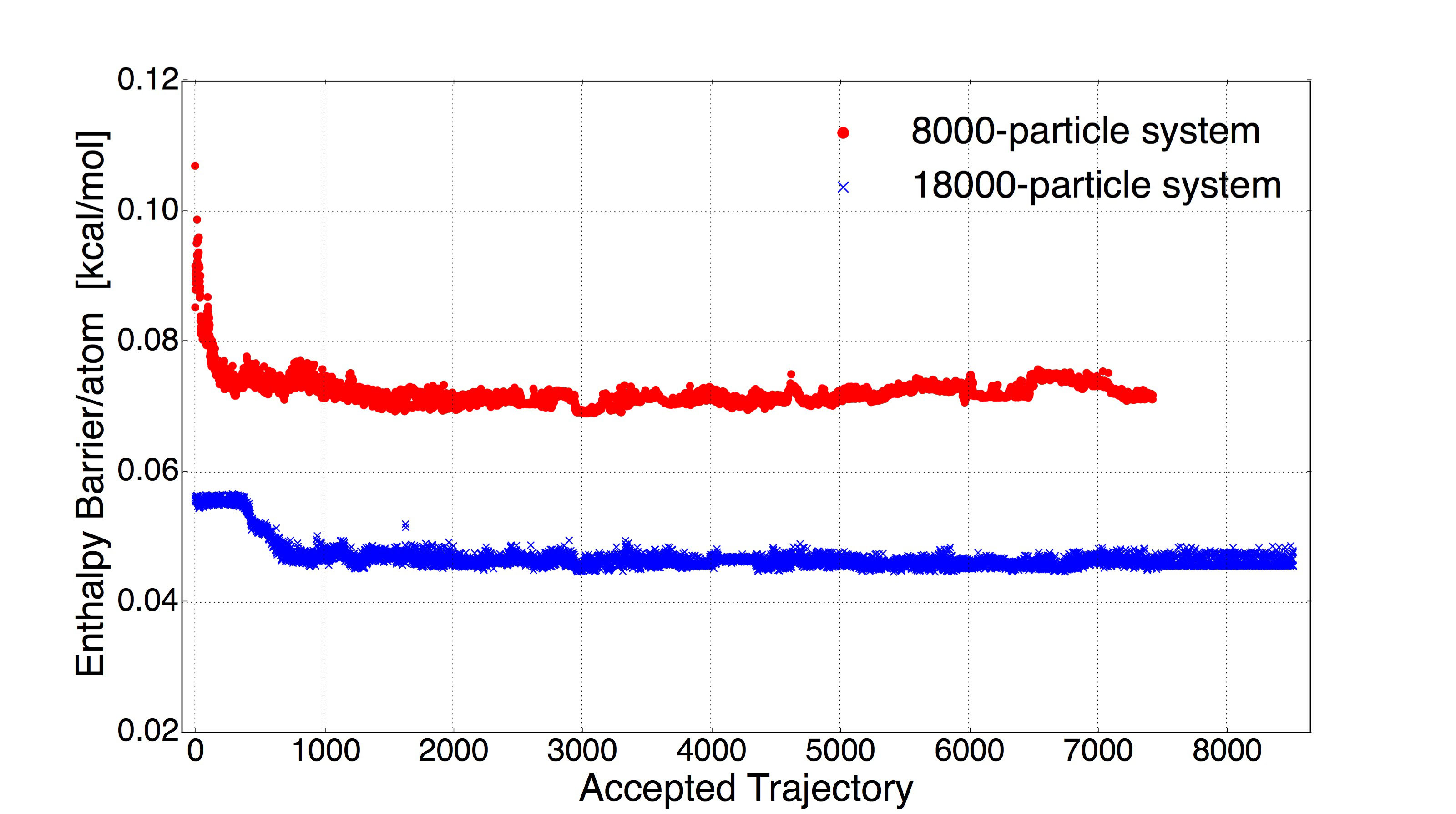}
\caption{\label{fig:18energy}Enthalpy barriers for 8000-- and 18000--particle system as a function of successful TPS trajectories. Not all the initial energies are shown due to limitations on the vertical scale. The enthalpy barrier per atom of the first trajectory in 8000--particle and 18000--particle TPS simulations are 0.849 kcal/mol and 0.533 kcal/mol respectively.}
\end{figure}

\subsection{Large System}

\begin{figure}[htbp!]
\captionsetup{justification=centering}
\includegraphics[width=0.5\textwidth]{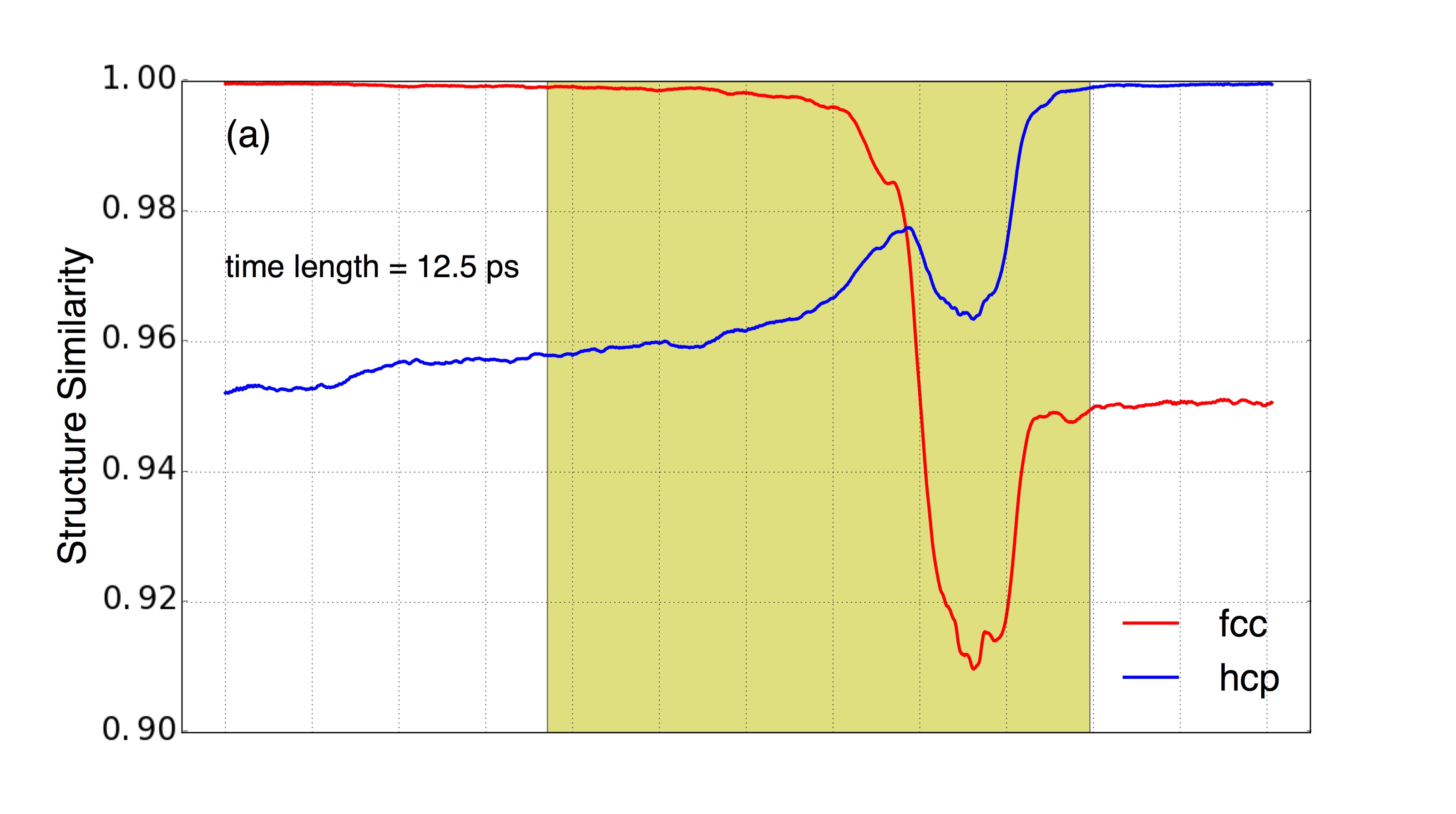}\\
\includegraphics[width=0.5\textwidth]{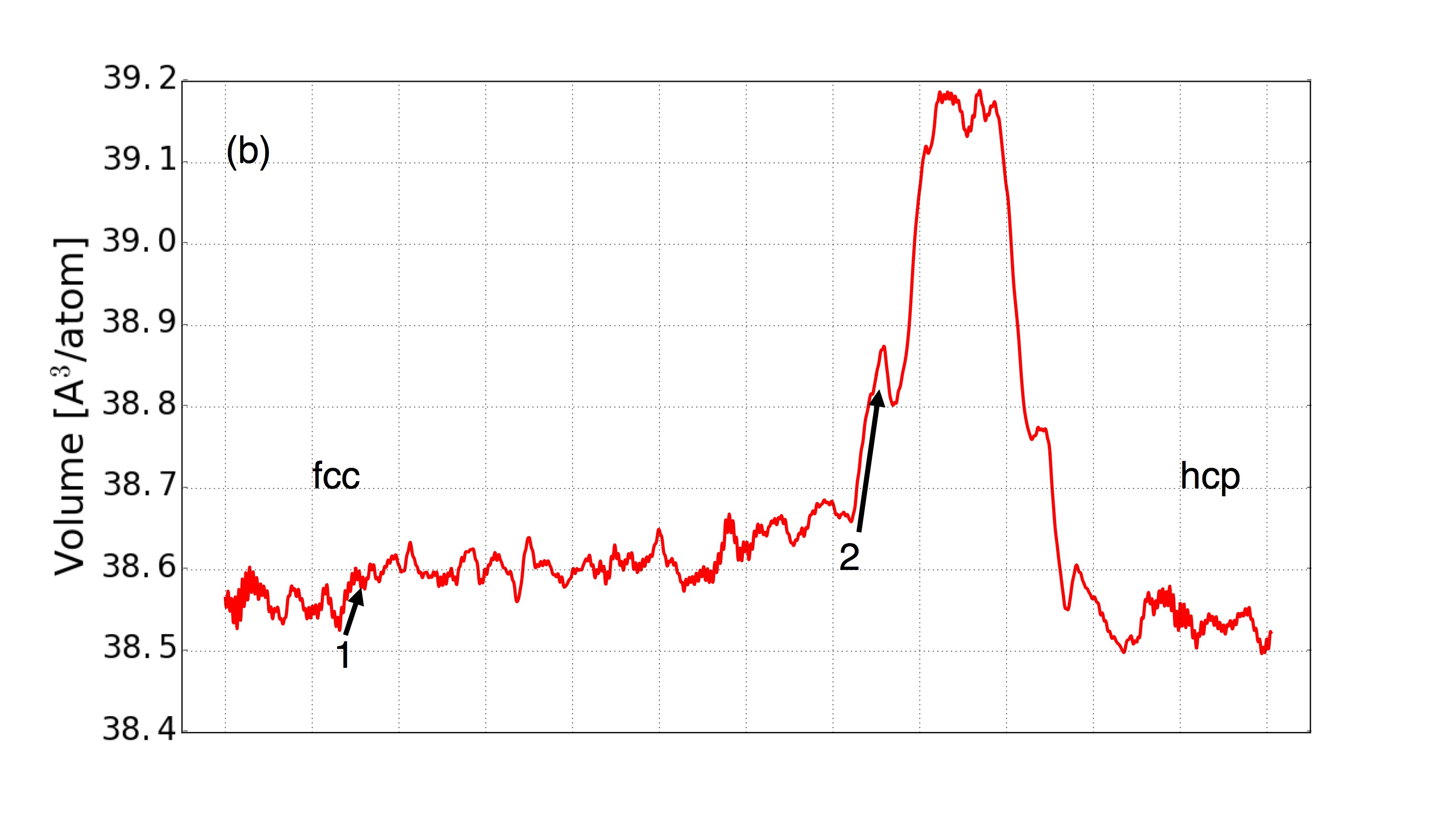}\\
\includegraphics[width=0.5\textwidth]{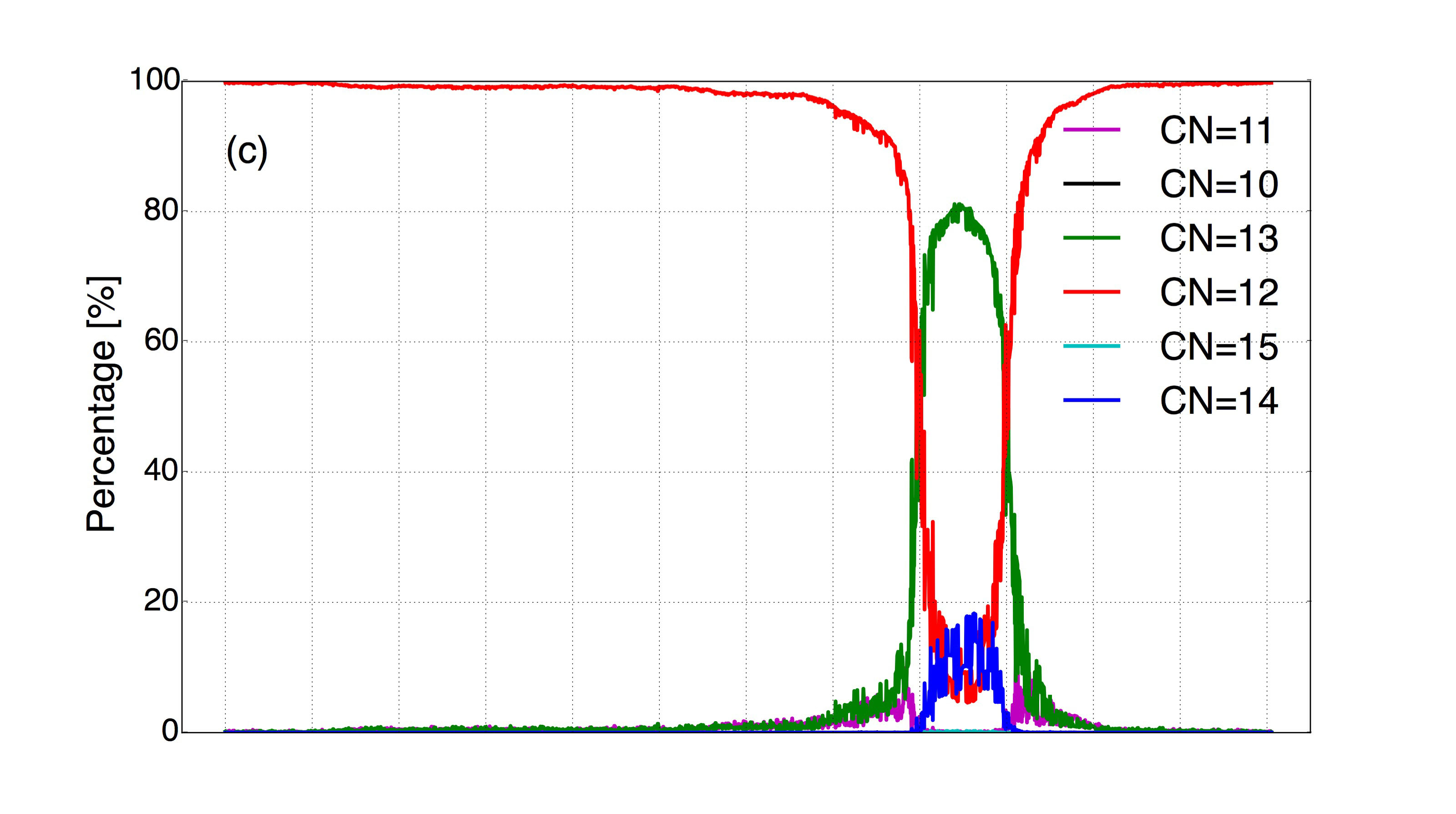}\\
\includegraphics[width=0.5\textwidth]{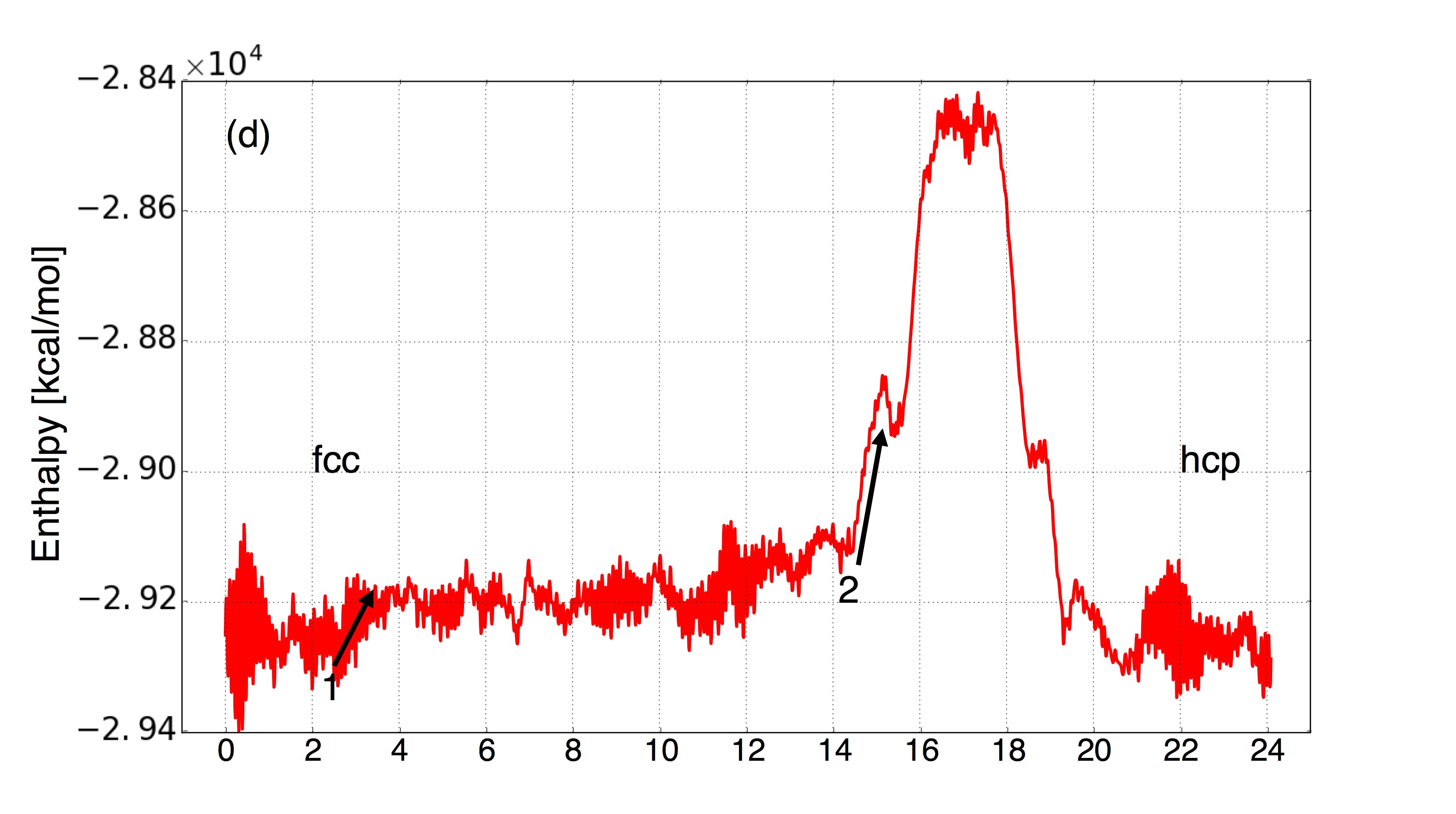}
\caption{\label{fig:18scv}A representative trajectory in 18000--particle system shows (a) structure similarities to {\it fcc} and {\it hcp} phases are plotted in red and blue, (b) atomic volume (c) percentages of atoms with different coordination numbers and (d) enthalpy during {\it fcc}--to--{\it hcp} transition.}
\end{figure}

\begin{figure*}[htbp!]
\captionsetup{justification=centering}
\includegraphics[width=\linewidth]{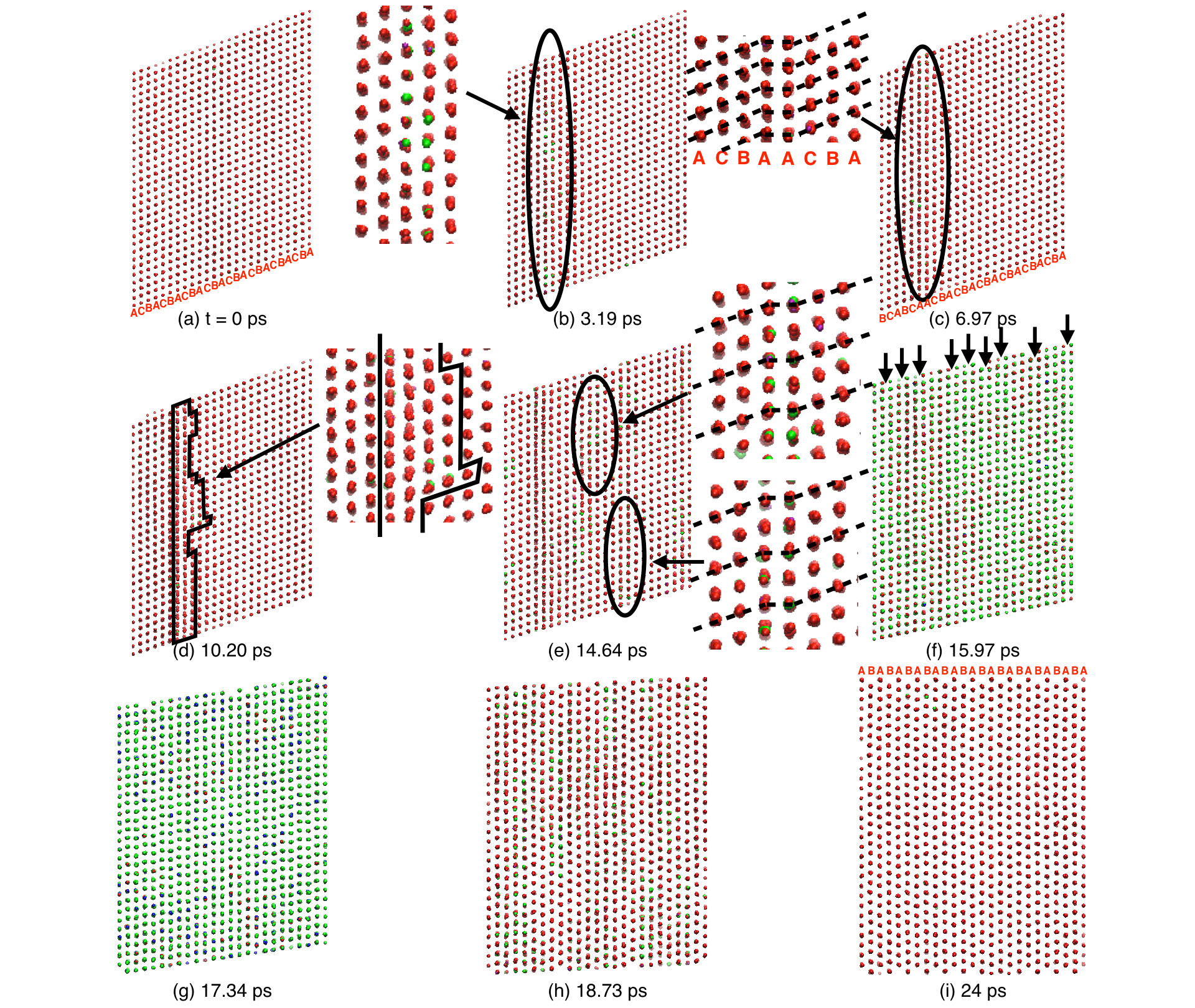}
\caption{\label{fig:transition18K}Snapshots of configurations during a transition in 18000--particle system, different colors are different coordination numbers. Atoms in red are 12--coordinated while green and blue are for 13-- and 14--coordinated atoms.}
\end{figure*}

As we did not observe nucleation in our 8000--particle system which is peculiar for a system undergoing a first order transition we performed simulations in a system with 18000--particles using the same analysis as before. The enthalpy barrier for the 18000--particle system turns out to be $826\pm 49$~kcal/mol. Here we find its enthalpy barrier per particle to be smaller than in the 8000--particle system (see Fig.~\ref{fig:18energy}) indicating that not the whole system is transitioning homogeneously at the same time, i.e. we see nucleation events. This is consistent with the findings  by inspecting individual transitions. The maximum deviation from both {\it hcp} and {\it fcc} structures in the fingerprint function is weaker than that in the smaller system. The structure similarity of a representative trajectory in large system is given in Fig.~\ref{fig:18scv}(a) as an example. The higher similarity to {\it hcp} and {\it fcc} structures of intermediate state shows that in contrast to the small system the large system does not completely loose the old structure before rebuilding the new one. Also the overall volume expansion during the transition is not as pronounced (Fig.~\ref{fig:18scv}(b)). It actually turns out that both, the 8000-- and the 18000--particle systems, have a very similar absolute maximal volume change. This suggests that the 8000 particle system discussed above is close to the correlated volume transitioning at the same time and that we need a large system to be able to have some decorrelation in space necessary for nucleation. In Fig.~\ref{fig:18scv}(c), we also find that with 18000 particles there is a small but not negligible portion of 12 coordinated atoms visible at all times in contrast to smaller system of 8000 particles above.

If we look at visualizations of a transition in Fig.~\ref{fig:transition18K}, it appears that the nucleation mentioned above is actually the formation and growth of stacking fault before the new phase grows in through the collective movement of atoms in the \{111\} planes. The system has the stacking sequence of \ldots{\it ABCABC}\ldots along the $\langle 111 \rangle$ direction at the beginning of the transition. After 3.19~ps, some defects are created locally in the system and they evolve into a stacking fault. The stacking sequence changes from \ldots{\it ABCABCABC}\ldots to \ldots{\it ABCAABCAB}\ldots in the circled region at 6.97~ps. During the formation of this stacking fault, we notice increases in both enthalpy and atomic volume, which are marked by arrowhead 1 in FIG.~\ref{fig:18scv}(b) and (c). Disorder is then created at one side of the stacking fault and its propagation yields another two stacking faults across the system at 14.64~ps. More stacking faults are generated before the collective movement of atoms in the whole system starts. They are marked with arrowheads in FIG.~\ref{fig:transition18K}(f). The growth of stacking faults is also indicated by rapid increases in enthalpy and atomic volume (arrowhead 2 in FIG.~\ref{fig:18scv}(b) and (c)). With these stacking faults, the transition is then carried on by the sliding of \{111\} planes, which involves the cooperative motions of atoms. In the most deformed state, more than 90\% atoms become over--coordinated. The system loses its coordinating characteristics as any crystalline solid. The 12--coordinates are regained for 90\% atoms at 18.73~ps, after which the system gradually relaxes into {\it hcp} state.

Except for the observation of stacking faults, it is noteworthy that we do not find qualitatively different transition types in the larger system. There are only quantitative differences in e.g. barrier heights. The transition in the 18000--particle system experiences the formation and growth of stacking faults over the system. The stacking faults facilitate the collective movement of atoms which is significant to the sliding of planes in this transition. Despite the assistance of stacking faults, the enthalpy barrier is still far beyond the thermal fluctuations at 40~K.

\section{Conclusions}
Our results show that {\it fcc}--to--{\it hcp} transformation in solid Ar at 40~K under ambient pressure is completed through the collective sliding of \{111\} planes in {\it fcc} structure which generates a high defect density. The system needs a large coherent volume for this transition such that only in our large system we find coexistence. The stacking sequence experiences disorder during its transition from \ldots{\it ABCABC}\ldots in {\it fcc} to \ldots{\it ABABAB}\ldots in {\it hcp}. Unlike the transition in Ar clusters~\cite{krainyukova2012observation}, no mechanism through an orthorhombic intermediate is observed. Instead we observe an intermediate state of lower order formed by 13-- and 14--coordinated atoms in all three types of transition of the small system; also in the large system there is no obvious {\it fcc}--{\it hcp} coexistence but rather a formation of a less crystalline solid, which in its order is between an amorphous system and a crystalline solid. The inaccessibility of {\it fcc}--to--{\it hcp} transformation in bulk Ar under low pressure and its sluggishness under high pressure~\cite{errandonea2002crystal,errandonea2002phase,errandonea2006structural,jephcoat1987pressure,PhysRevB.77.052101,freiman2008raman,PhysRevB.79.132101,verkhovtseva2003atomic,danylchenko2004electron,danil2008electron,krainyukova2012observation,danylchenko2014electron} can be explained by our results on the transition enthalpy barrier and the large correlated volume, which is far beyond the thermal fluctuation.

All transition pathways in the small system are classified into three types in terms of lattice deformation time and relaxation time. But they all have similar amorphous intermediates during the phase transition. The formation and growth of stacking faults in large system do not change the transition mechanism analyzed from small system but facilitate the transition by creating abundant local defects.

The TPS method employed in our research provides an atomistic understanding of the Ar phase transition. Our results explain the inaccessibility and sluggishness of this transition in experiments and predicts a different intermediate.

It is possible that with even larger systems the transition localizes further. It would be interesting to perform TPS over even larger system in order to explore the nucleation event. Challenges lie in addition to the point of raw computer power in the initialization of the first trajectory for TPS in larger systems and more efficient sampling of different transition regions in the transition path ensemble.


\bibliography{aip1002}

\clearpage
\appendix

\section*{Supplementary Data}
\section{Computation of Coordination Number}

The coordination number is computed as the number of neighbor atoms within a specified cutoff distance from the central atom. The first minimums of the Radial Distribution Function (RDF) defines the cutoff distance to find nearest neighbors of central atoms in a given structure. By computing the RDFs, we determine the first minimum locations of individual structures in a representative transition of 18000--particles and three types transitions of 8000-particles.

\begin{figure}[h!]
\captionsetup{justification=centering}
\includegraphics[width=0.5\textwidth]{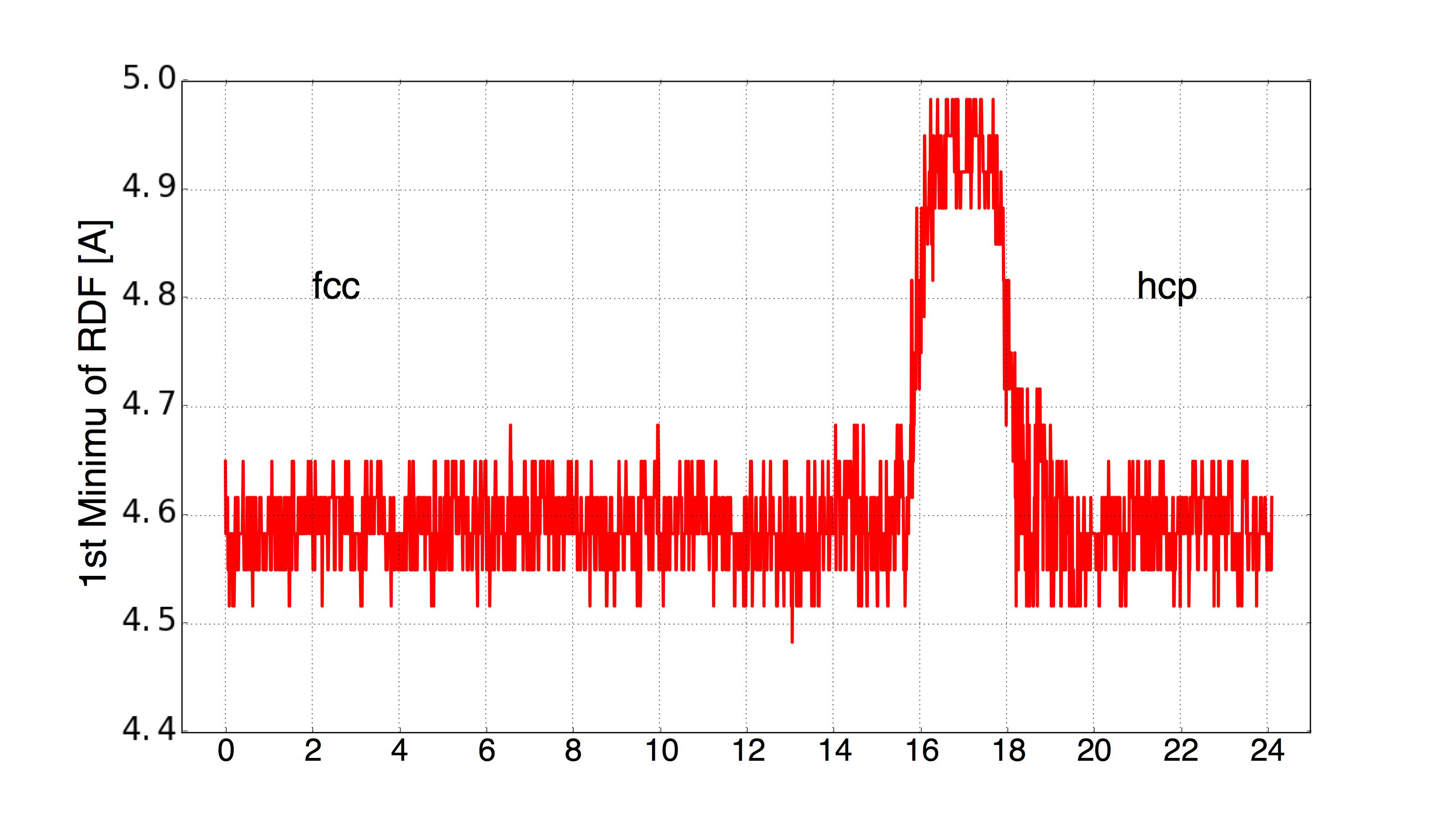}
\caption{\label{fig:18Kmin}First minimum position of structures during a  a representative transition in 18000--particle system.}
\end{figure}

The FIG.~\ref{fig:18Kmin} gives the RDF first minimums of transitory structures along a representative transition in 18000--particle system. There is a 8.7\% radial expansion for the most deformed state in FIG. 7(a). The increase in atomic spacing during the transition explains the volume expansion in spite of the over--coordinates of atoms.

\begin{figure}[h!]
\captionsetup{justification=centering}
\includegraphics[width=0.5\textwidth]{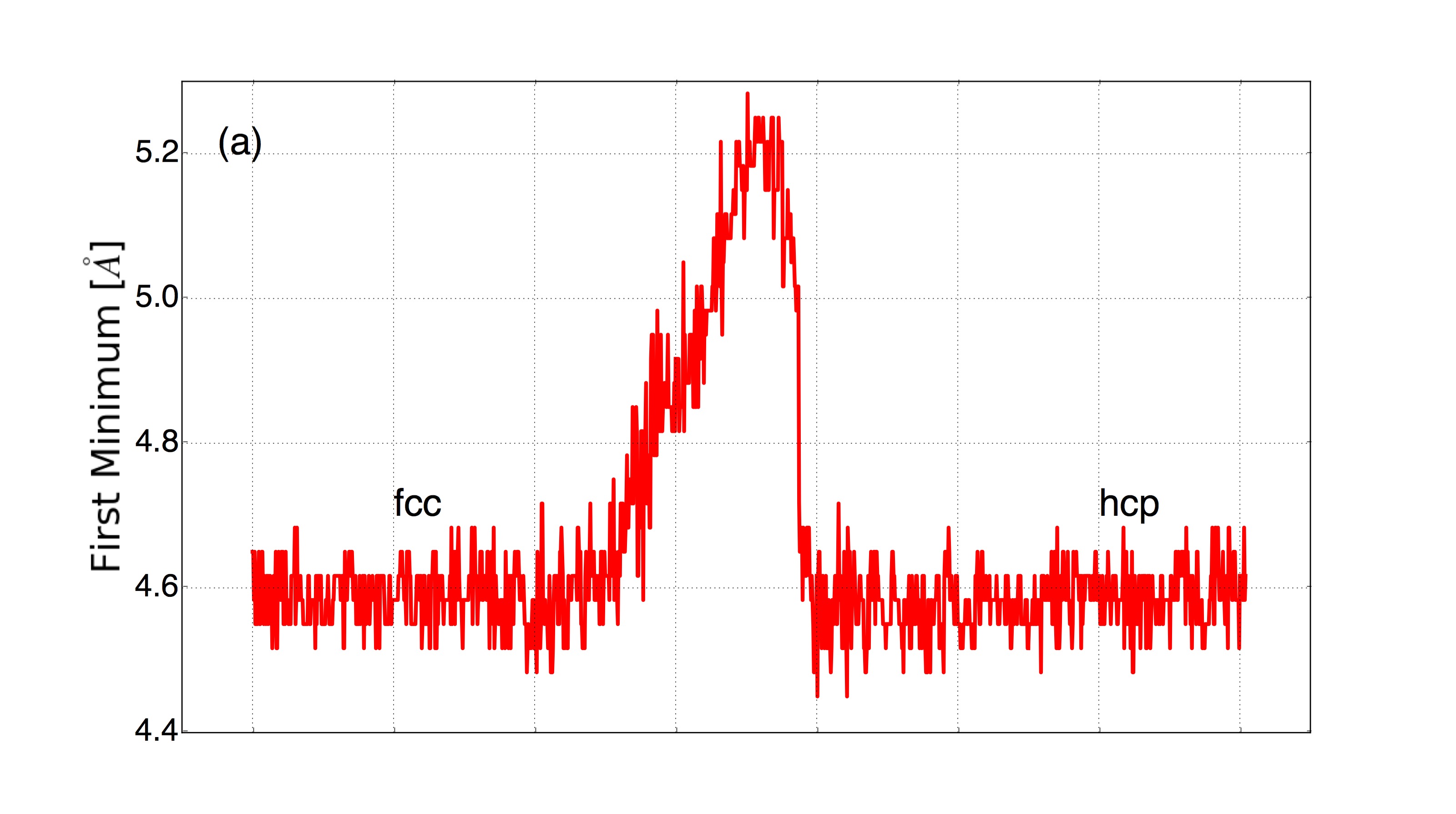}
\includegraphics[width=0.5\textwidth]{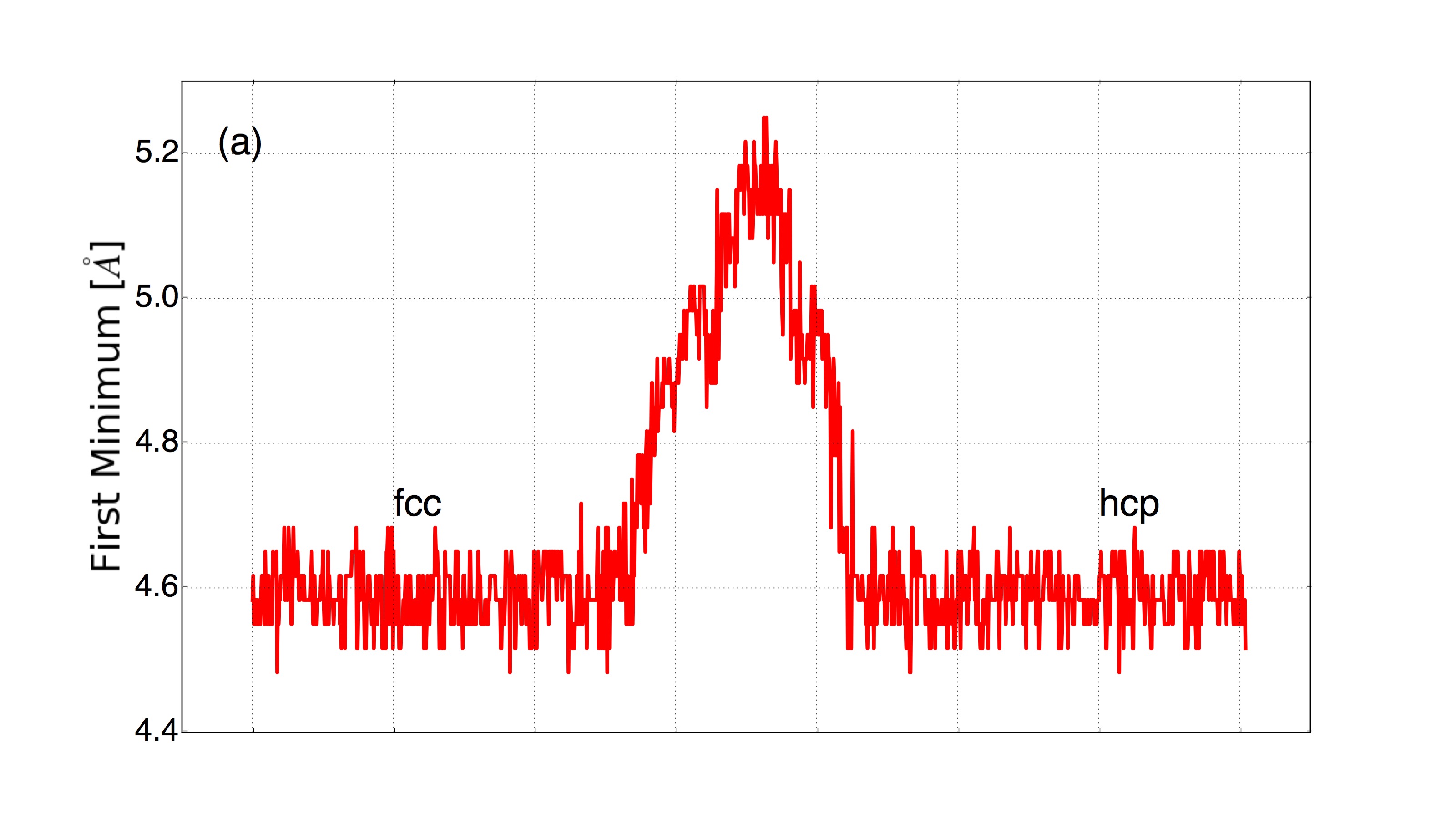}
\includegraphics[width=0.5\textwidth]{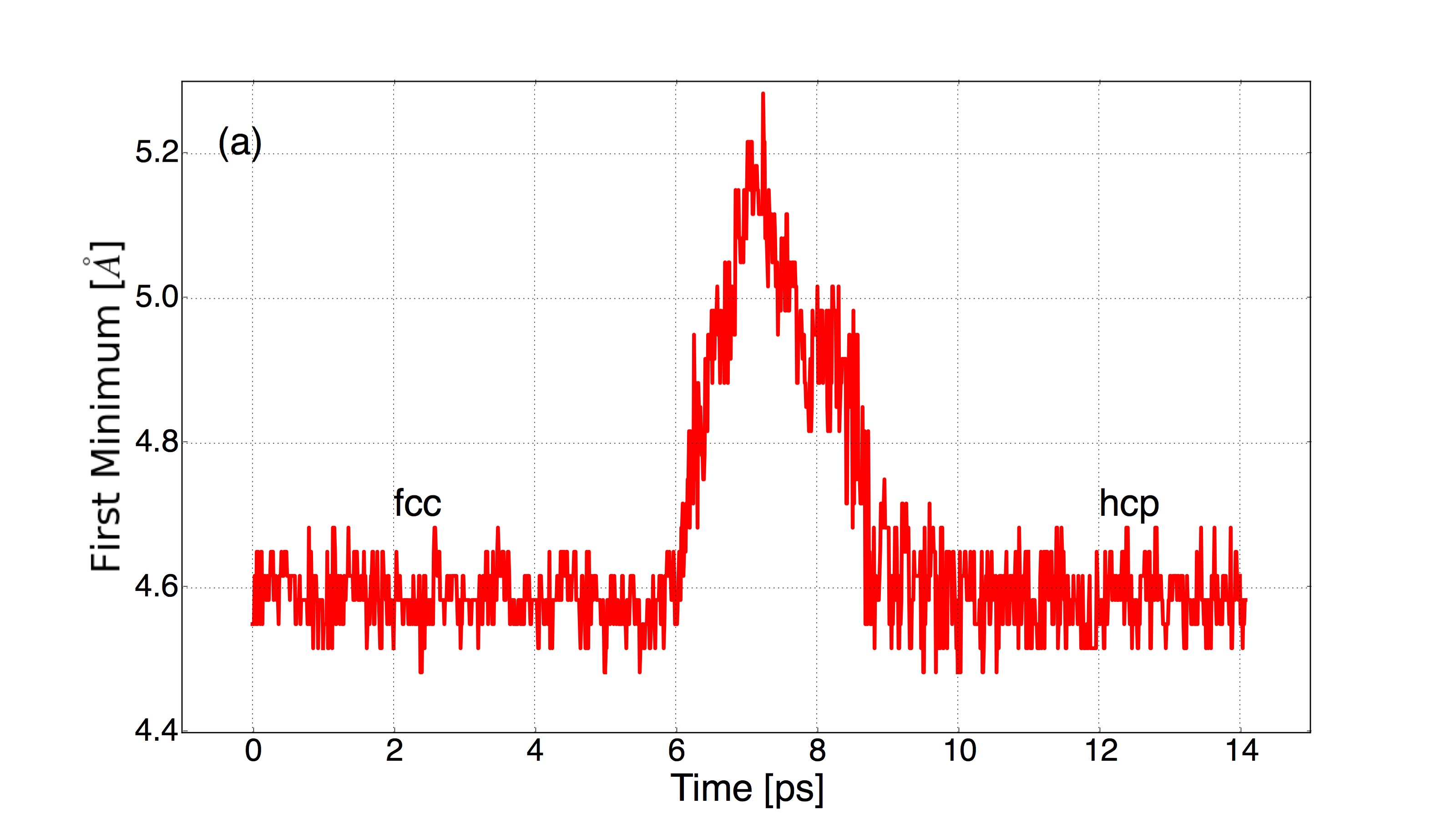}
\caption{\label{fig:rdf} First minimum position of structures during the transition of (a) Trajectory \uppercase\expandafter{\romannumeral1}, (b) Trajectory \uppercase\expandafter{\romannumeral2} and (c) Trajectory \uppercase\expandafter{\romannumeral3}.}
\end{figure}

The FIG.~\ref{fig:rdf} gives the RDF first minimums of transitory structures during three types of transition \uppercase\expandafter{\romannumeral1} to \uppercase\expandafter{\romannumeral3} in 8000--particle system. An approximate 14\% radial expansion happens to all the most deformed states in FIG. 1--3(b) during these transitions, which leads to the volume expansion.

In both FIG.~\ref{fig:18Kmin} and FIG.~\ref{fig:rdf}, the \textit{fcc} and \textit{hcp} structures have their RDFs' first minimums at around 4.6~\AA. Compared to enthalpy profiles of these transitions in paper(FIG. 7(d) FIG. 1--3(a)), the first minimums profiles of these transitions shows great consistency, which indicates the major enthalpy contribution of nearest neighbors to central atoms.

We then compute coordination numbers of atoms within these first minimums distances for every structures in each transition. The ratios of different coordinates in every structures of these transitions are therefore gained in FIG.~7(c) and FIG.~1--3(c).

\section{Volume Changes during Phase Transition}

Here we use Type uppercase\expandafter{\romannumeral1} transition in 8000--particle system as an example to analyze the consistencies of volume change with enthalpy, structure similarity and coordinates.

FIG.~4 in the main text demonstrates shrinkage and expansion along two axes. This phenomenon is evidenced by our measurement of the volume of cell during phase transition. According to FIG.~5 in paper, the cell expands and shrinks during the \textit{fcc}-\textit{hcp} transformation with maximum expansion being 2.59\%. This change in volume lead to corresponding changes along axes shown in FIG.~4.

Further comparisons of volume with energetic and structural profiles in FIG.~1(a) and (b) show remarkable consistencies. The cell begins to expand since the cell loses its similarity to \textit{fcc} phase. The shrinkage of cell happens when the cell relaxes to \textit{hcp} phase. The volume change also induces a variation of the average interatomic distances between atoms, which result in simultaneous increases and decreases in enthalpy. By looking into the coordinating environment of atoms, we find that the volume change can be well explained by the generation and dissipation of defects through the phase transition. Since the growth of defects, the \textit{fcc} cell loses its close packed structure and its volume therefore increases. The cell then shrinks until the dense arrangement of atoms is regained in \textit{hcp} phase. According to Fig~1(c), the percentage of over--coordinated atoms (CN = 13, 14) is higher than that of under--coordinated atoms (CN = 10, 11). As at the same time the density decreases it appears that the increased mobility leads to this effect.

\section{Crystallinity of Intermediate State}
In this section, the crystallinity of intermediate states in two systems are compared to random, amorphous, \textit{fcc} and \textit{hcp} respectively by using structure function, $r\cdot g(r)$, where $g(r)$ is the radial distribution function and $r$ is the distance from the center atom. The random solid has randomly distributed atoms and its structures function shows a simple linear behavior. The amorphous solid lacks long--range order. Its structure function is similar to the random solid in long distance. \textit{fcc} and \textit{hcp} are both ordered structures and have crystalline characteristics in the short and long range regions. Therefore we see sharp peaks over a wide range of distances in their structure function. The breadth and height of the peaks in the structure function are indicative of the crystallinity of the solid. A qualitative understanding of intermediates' crystallinity can be gained by comparing structure function to these solids.

\subsection{Small System}
For the transition in the system of 8000 particles, the intermediate state has a similar structure function as amorphous solid. However, the peaks of the intermediate state are sharper in the long range. Compared to \textit{fcc} and \textit{hcp}, the intermediate state has broader and less sharp peaks. Therefore the order of intermediate state in small system is between amorphous and crystalline solids.

\begin{figure}[htbp!]
\captionsetup{justification=centering}
\includegraphics[width=0.5\textwidth]{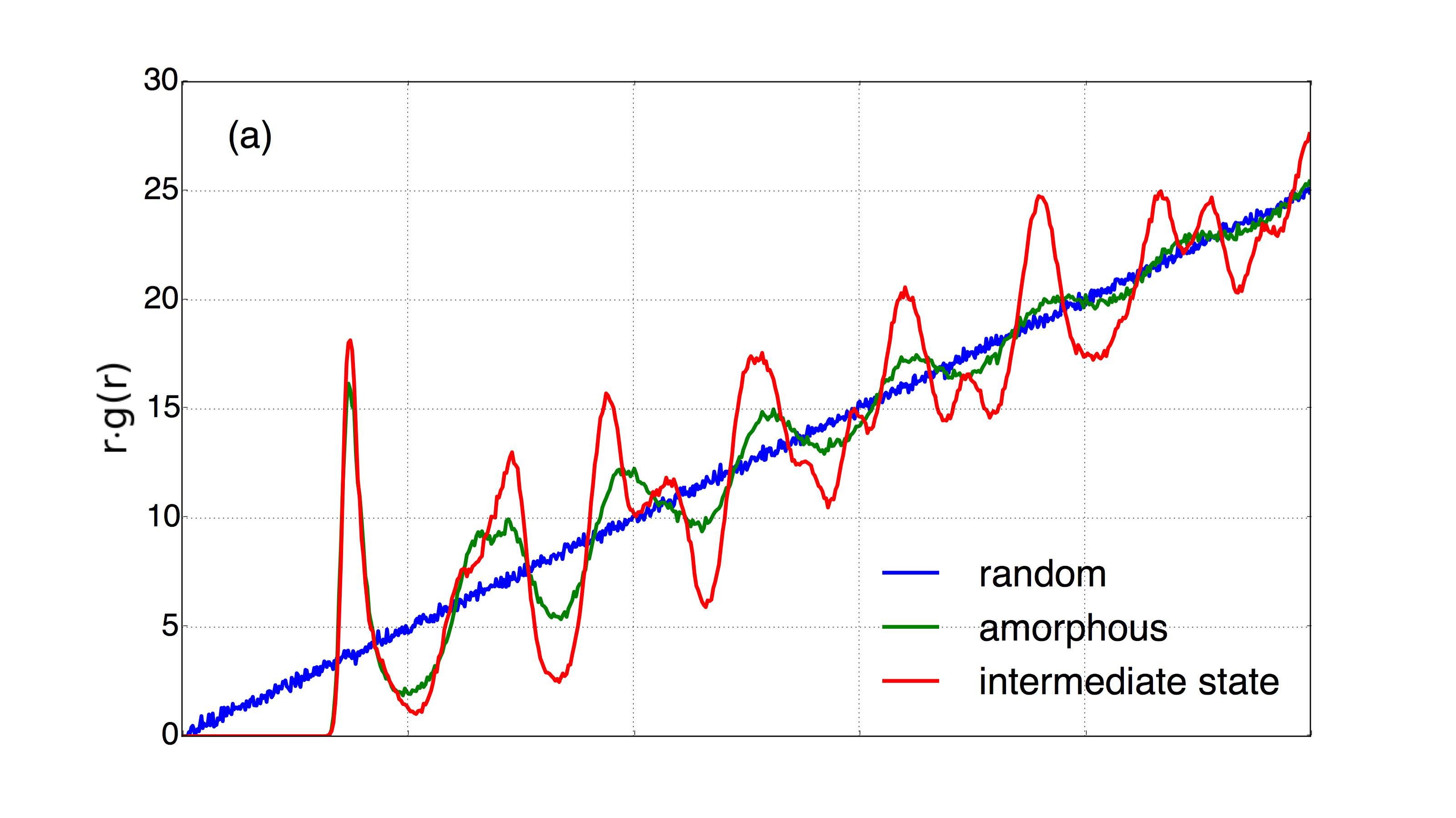}
\includegraphics[width=0.5\textwidth]{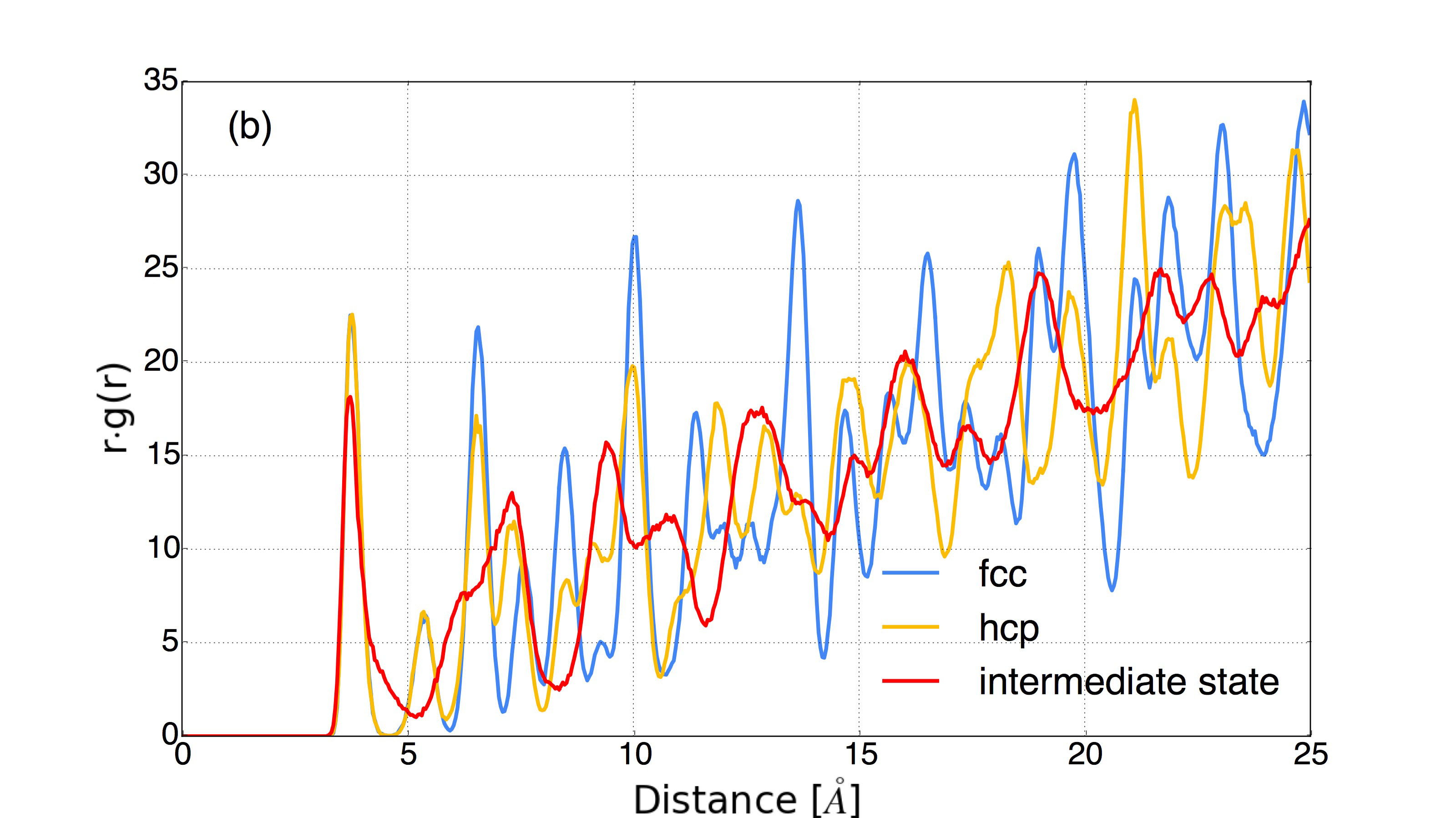}
\caption{\label{fig:8kStrucFunc} Structure Function of (a) random, amorphous and intermediate solid (b) fcc and hcp solid in system of 8000 particles.}
\end{figure}

\subsection{Large System}
For the transition in the system of 18000 particles, the intermediate state has more sharp features than the structure function of the amorphous solid. The shape of the intermediate state structure function is very similar to the \textit{hcp} state and this is evidenced by more than 96\% structure similarity between them shown in the FIG. 7(a) of manuscript. These similarities are gained through the local growth of stacking faults in large system, which facilitates the collective planar movements of atoms and reduces the structure deformation.

Despite similarities to a crystalline solid, this intermediate state has lower and broader peaks in the structure function when compared to \textit{hcp}. It is also noticeable that the volume expansion as discussed in FIG. 7(b) leads to a shifting of the intermediate structure function. So the order of intermediate state in large system is still between amorphous and crystalline solids.

\begin{figure}[htbp!]
\captionsetup{justification=centering}
\includegraphics[width=0.5\textwidth]{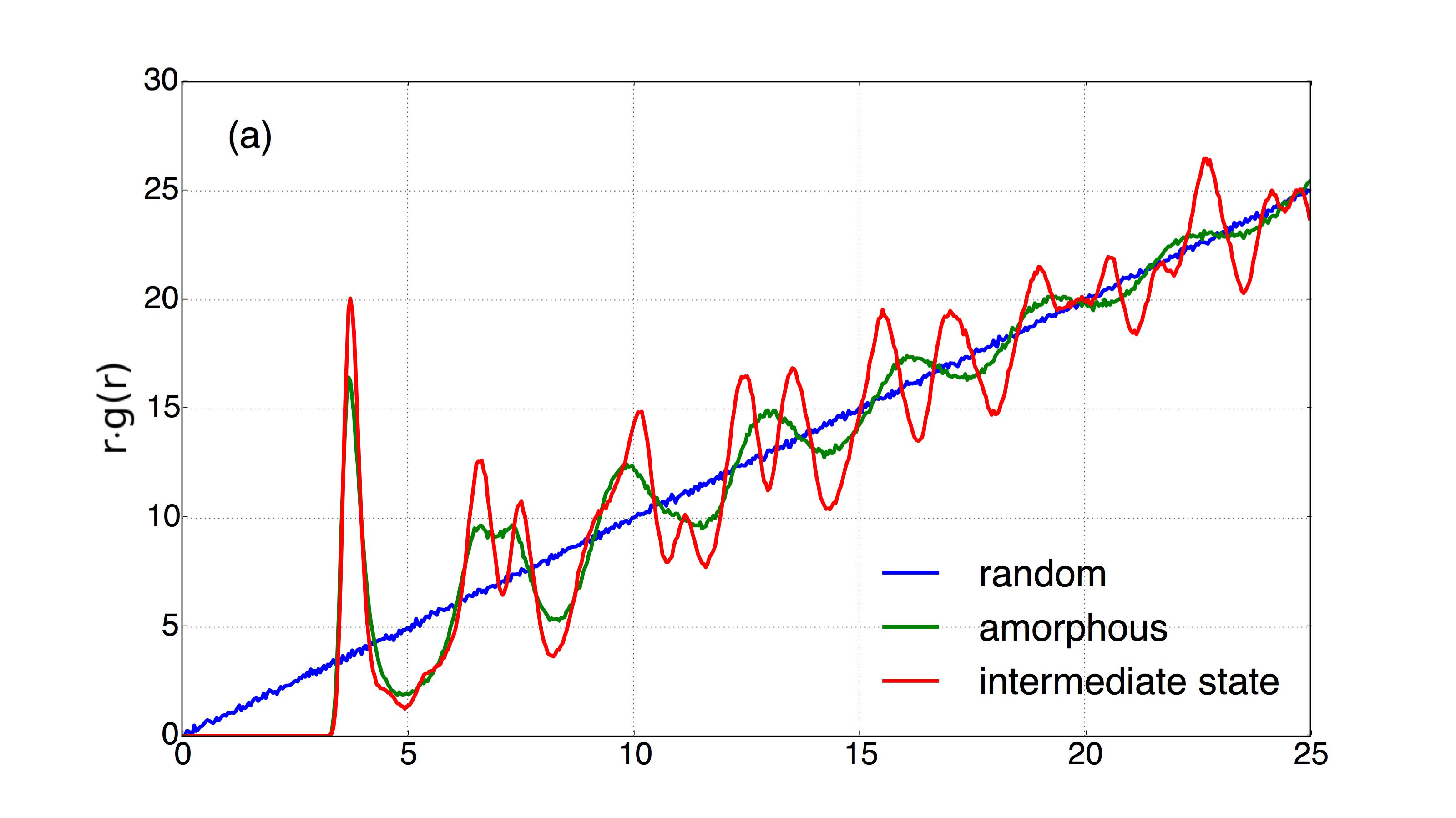}
\includegraphics[width=0.5\textwidth]{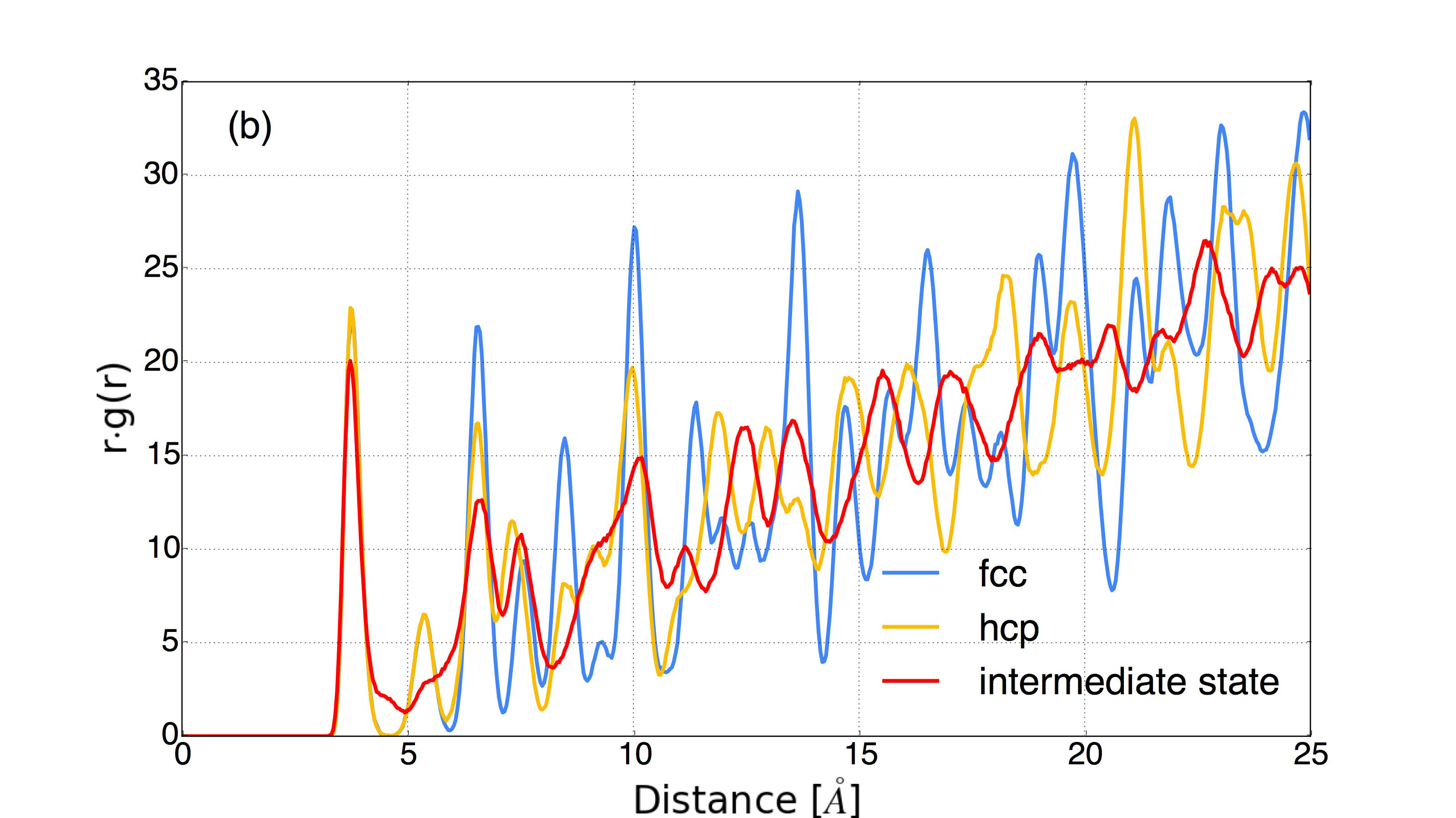}
\caption{\label{fig:18kStrucFunc} Structure Function of (a) random, amorphous and intermediate solid (b) fcc and hcp solid in system of 18000 particles.}
\end{figure}

\end{document}